\documentclass[usletter]{article}

\usepackage{amsmath}
\usepackage{amsthm}
\usepackage{amssymb}
\usepackage{times}
\usepackage{tikz}
\usepackage{enumitem}
\usepackage{url}

\usepackage[totalwidth=450pt,totalheight=640pt]{geometry}

\newtheorem{proposition}{Proposition}
\newtheorem{lemma}{Lemma}
\newtheorem{theorem}{Theorem}
\newtheorem{corollary}{Corollary}

\usepackage{amsmath}
\usepackage{amssymb}
\usepackage{times}
\usepackage{tikz}
\usepackage{enumitem}
\usepackage{url}

\newcommand{\hy}{\hbox{-}\nobreak\hskip0pt}

\newcommand{\CCC}{\mathcal{C}}

\newcommand{\cwd}{\text{\normalfont cw}}
\newcommand{\hw}{\text{\normalfont hw}}
\newcommand{\bhw}{\beta\text{-\normalfont hw}}

\newcommand{\SB}{\{\,}
\newcommand{\SM}{\;{|}\;}
\newcommand{\SE}{\,\}}
\newcommand{\Card}[1]{|#1|}

\newcommand{\SSS}{\mathcal{S}}
\newcommand{\TTT}{\mathcal{T}}

\newcommand{\CNF}{{\sf CNF}}
\newcommand{\BAC}{{\sf BAC}}
\newcommand{\DPS}{{\sf DPS}}

\newcommand{\DP}{\text{\normalfont DP}}
\renewcommand{\P}{{\sf P}}
\newcommand{\NP}{{\sf NP}}

\newcommand{\FPT}{\text{\normalfont FPT}}
\newcommand{\XP}{\text{\normalfont XP}}
\newcommand{\W}[1][xxxx]{\text{\normalfont W[#1]}}
\let\phi=\varphi

\newcommand{\var}{\mbox{var}}

\newcommand{\ol}[1]{\overline{#1}}

\newcommand{\acyc}{\text{\sc -Acyc}}

\newcommand{\FALL}{F_a}
\newcommand{\FTWO}{F_s}
\newcommand{\FCYC}{F_c}
\newcommand{\FALLCYC}{F_{ac}}

\begin{document}

\title{Satisfiability of Acyclic and Almost Acyclic CNF
  Formulas\thanks{This is the author's self-archived copy of a paper
that    appeared in \emph{Theoretical Computer Science},
vol. 481, pp. 85-99, 2013. Extended abstracts appeared in the Proceedings of FSTTCS 2010 and SAT 2011.
}}

\author{Sebastian Ordyniak\thanks{Supported by
ERC (COMPLEX REASON, 239962).}\\
  \small  Masaryk University, Brno, Czech Republic
 \and Daniel Paulusma\thanks{Supported by EPSRC (EP/G043434/1).}\\
  \small  Durham University, Durham, UK 
 \and Stefan Szeider${}^\dagger$\\
  \small Vienna University of Technology, Vienna, Austria
}

\date{}

\maketitle

\begin{abstract}
We show that the {\sc Satisfiability} (SAT) problem for CNF formulas with $\beta$-acyclic hypergraphs can be solved in polynomial time
by using a special type of Davis-Putnam resolution where each resolvent
is a subset of a parent clause. 
We extend this class to CNF formulas for which this type of Davis-Putnam resolution still applies and
show that testing membership in this class is \NP-complete.
We compare the class of $\beta$-acyclic formulas 
and this superclass with a number of known polynomial
formula classes.
We then study the parameterized complexity of SAT for ``almost'' $\beta$-acyclic instances, using as parameter the formula's distance
from being $\beta$-acyclic. As distance we use the size of smallest
strong backdoor sets and the $\beta$-hypertree width.  As a by-product
we obtain the $\W[1]$-hardness of SAT parameterized by the
(undirected) clique-width of the incidence graph, which disproves a
conjecture by Fischer, Makowsky, and Ravve.

\medskip
\noindent
{\bf Keywords} acyclic hypergraph, chordal bipartite graph, 
Davis-Putnam resolution. 
\end{abstract}


\maketitle

\section{Introduction}

We consider the 
{\sc Satisfiability} (SAT) problem on classes of CNF formulas (formulas in Conjunctive Normal Form) with restrictions on their associated 
hypergraphs, which are obtained from these formulas by
ignoring negations and considering clauses as hyperedges on variables.
This is a natural study,
because many computationally hard problems can be solved efficiently on acyclic instances.
However, there are several notions of acyclicity for
hypergraphs:
$\alpha$-acyclicity, $\beta$-acyclicity,
$\gamma$-acyclicity, and Berge acyclicity.
We  provide the relevant definitions in Section~\ref{sec:pre} and
refer to Fagin~\cite{Fagin83} for a detailed description.
The notions of acyclicity
are strictly ordered with respect to their
generality: 
\begin{equation}
  \label{eq:chain}
  \alpha\acyc \supsetneq \beta\acyc \supsetneq \gamma\acyc \supsetneq
  \text{Berge}\acyc
\end{equation}
where $X\acyc$ denotes the class of $X$-acyclic hypergraphs, which are in 1-to-1 correspondence to a class of CNF formulas called $X$-acyclic formulas.
It is known that SAT is \NP-complete for 
$\alpha$-acyclic formulas~\cite{SamerSzeider10} and
polynomial-time solvable for  
Berge-acyclic formulas~\cite{FischerMakowskyRavve06,SamerSzeider10}.

\paragraph{Our Results}
In Section~\ref{sec:poly} we determine the 
boundary between \NP-completeness and polynomial-time tractability 
in the chain  (\ref{eq:chain}) by showing that SAT is
polynomial-time solvable for $\beta$-acyclic formulas.
Consequently, the same holds for $\gamma$-acyclic formulas.
To prove our result we use 
a fundamental procedure called the Davis-Putnam procedure,
which successively eliminates variables using 
Davis-Putnam resolution~\cite{DavisPutnam60}. In general, this procedure 
is not efficient, because the number of clauses may
increase after each application of Davis-Putnam resolution.
However, $\beta$-acyclic formulas are related to chordal 
bipartite graphs~\cite{TarjanYannakakis84}, and this allows us to compute
an elimination ordering of the variables with the 
property that each obtained resolvent is a subset of a parent
clause. This type of resolution is known as subsumption
  resolution~\cite{KL99}.  
  
In Section~\ref{s-gen} we show that there
are CNF formulas that are not $\beta$-acyclic but that still admit an
elimination ordering of their
variables based on subsumption resolution, such that the Davis-Putnam
procedure takes polynomial time.  We call such an elimination ordering
\emph{DP-simplicial}.  This leads to a new class $\DPS$ of CNF formulas
that contains the class of $\beta$\hy acyclic formulas.
In
Section~\ref{s-npcomplete} we show that testing membership in this
class is an \NP-complete problem.  The reason for the \NP-hardness is
that a formula may have several so-called \emph{DP-simplicial}
variables, one of which must be chosen to be eliminated but we do not
know which one.  
In Section~\ref{s-inter} we show how to work around
this obstacle to some extent, i.e., we identify a subclass of $\DPS$
that is a proper superclass of the class of $\beta$-acyclic formulas for
which SAT is polynomial-time solvable.
In Section~\ref{s-comp} we show that the class of
$\beta$\hy acyclic formulas 
and its superclass $\DPS$ are \emph{incomparable}
with other known polynomial classes of CNF
formulas.  Hence, $\beta$\hy acyclic formulas
form a new ``island of
tractability'' for SAT.

In Section~\ref{s-parameterized}  we 
study the complexity of SAT for formulas that are parameterized by their ``distance'' from the
class of $\beta$-acyclic CNF formulas. We use two distance
measures.
The first distance measure is based on the notion of a \emph{strong
  backdoor set}. For a CNF formula $F$ we define its ``distance to
$\beta$-acyclicity'' as the size $k$ of a smallest set $B$ of variables
such that for each partial truth assignment to $B$, the reduct of $F$
under the assignment is $\beta$-acyclic; such a set $B$ is a strong
backdoor set.  If we know $B$, then deciding the satisfiability
of $F$ reduces to deciding the satisfiability of at most $2^k$
$\beta$-acyclic CNF formulas, and is thus fixed-parameter tractable with
respect to~$k$. We show, however, that finding such a set $B$ of size
$k$ (if it exists) is $\W[2]$-hard, thus unlikely fixed-parameter
tractable for parameter $k$, 
which limits the algorithmic usefulness of this distance measure.

The second distance measure we consider is the \emph{$\beta$-hypertree
  width}, a hypergraph invariant introduced by Gottlob and
Pichler~\cite{GottlobPichler04}.  The classes of hypergraphs of
$\beta$-hypertree width $k=1,2,3,\dots$ form an infinite chain of proper
inclusions. Hypergraphs of $\beta$-hypertree width~1 are exactly the
$\beta$-acyclic hypergraphs. Thus $\beta$-hypertree width is also a way
to define a ``distance to $\beta$-acyclicity.''  The complexity of
determining the $\beta$-hypertree width of a hypergraph is
open~\cite{GottlobPichler04}. However, we show that SAT  
parameterized by an upper bound on the $\beta$-hypertree width is
$\W[1]$-hard even if we are given
the CNF formula together with a 
$\beta$-hypertree decomposition of width $k$.
As a side effect, we obtain from this result that SAT is
also $\W[1]$-hard when parameterized by the \emph{clique-width} (of the
undirected incidence graph) of the CNF formula. 
This disproves a
conjecture by Fischer, Makowsky, and
Ravve~\cite{FischerMakowskyRavve06}.

\section{Preliminaries}\label{sec:pre}

In this section we state our basic terminology and notations. We also present some known results that will be useful at several places in the paper.

\subsection{Formulas and Resolution}\label{s-formulas}

We assume an infinite supply of propositional \emph{variables}. A
\emph{literal} is a variable $x$ or a negated variable $\ol{x}$; if
$y=\ol{x}$ is a literal, then we write $\ol{y}=x$.  For a set $S$ of
literals we put $\ol{S}=\SB\ol{x} \SM x\in S\SE$; $S$ is
\emph{tautological} if $S\cap \ol{S}\neq \emptyset$.  A \emph{clause} is
a finite non-tautological set of literals.  A finite set of clauses is a
\emph{CNF formula} (or \emph{formula}, for short).
A variable $x$ \emph{occurs} in a clause
$C$ if $x \in C\cup \ol{C}$; $\var(C)$ denotes the set of variables
which occur in $C$.  
A variable $x$ {\it occurs} in a formula $F$
if it occurs in one of its clauses, and 
we put $\var(F)=\bigcup_{C\in F} \var(C)$.  
If $F$ is a formula and $X$ a set of variables, then we denote by
$F-X$ the formula obtained from $F$ after removing all literals $x$
and $\ol{x}$ with $x \in B$ from the clauses in $F$. If $X=\{x\}$ we
simply write $F-x$ instead of $F-\{x\}$.

Let $F$ be a formula and $X \subseteq \var(F)$.
A \emph{truth assignment} is a mapping $\tau:X\rightarrow \SB 0,1 \SE$
defined on some set $X$ of variables; we write $\var(\tau)=X$.  For
$x\in \var(\tau)$ we define $\tau(\ol{x})=1-\tau(x)$.  
For a truth
assignment $\tau$ and a formula $F$, we define
\[
F[\tau]=\SB C\setminus \tau^{-1}(0)
\SM C\in F,\ 
  C\cap\tau^{-1}(1)= \emptyset\SE, 
\]
i.e., $F[\tau]$ denotes the result of instantiating variables according
to $\tau$ and applying the usual simplifications. 
A truth assignment $\tau$ \emph{satisfies} a clause $C$ if $C$ contains some literal $x$ with $\tau(x)=1$; $\tau$ satisfies a formula $F$ if it satisfies all
clauses of $F$.  
A formula is
\emph{satisfiable} if it is satisfied by some truth assignment; 
otherwise it is \emph{unsatisfiable}. 
Two formulas $F$ and $F'$ are
\emph{equisatisfiable} if either both are satisfiable or both are
unsatisfiable. 
The \textsc{Satisfiability} (SAT) problem asks
whether a given CNF formula is satisfiable.

Let $C,D$ be two clauses such that $C\cap \ol{D}=\{ x \}$ for a variable
$x$. 
The clause $(C \cup D) \setminus \{ x,\ol{x} \}$ is called the
\emph{$x$-resolvent} (or {\it resolvent}) of $C$ and $D$; the clauses
$C$ and $D$ are called \emph{parent clauses} of the $x$-resolvent. Note
that by definition any two clauses have at most one resolvent.  Let $F$
be a formula. A sequence $C_1,\dots,C_n$ is a \emph{resolution
  derivation} of $C_n$ from $F$ if every $C_i$ is either in $F$ or the
resolvent of two clauses $C_j$ and $C_{j'}$ for some $1\leq j<j'\leq
i-1$.
If $C_n$ is the empty clause, then the sequence is called a \emph{resolution
refutation} of~$F$.
The derivation is \emph{minimal} if we cannot
delete a clause from it and still have a resolution derivation of $C_n$
from $F$.  We call a clause $C_n$ a \emph{resolution descendant}
of a clause $C_1 \in F$ if there is a minimal resolution derivation
$C_1,\dots,C_n$ of $C_n$ from $F$.

Consider a formula $F$ and a variable $x$ of $F$.  
Let $\DP_x(F)$ denote the formula obtained from $F$ 
after adding all possible $x$-resolvents and removing all clauses in 
which $x$ occurs. We say that $\DP_x(F)$ is
obtained from $F$ by \emph{Davis-Putnam resolution}, 
and that we {\it eliminated} $x$.
It is well known (and easy to show) that $F$
and $\DP_x(F)$ are equisatisfiable.

For an ordered sequence
of variables $x_1,\dots,x_k$ of $F$, we set
$\DP_{x_1,\dots,x_k}(F)=\DP_{x_k}( \cdots (\DP_{x_1}(F)) \cdots )$ 
and $\DP_\emptyset(F)=F$.
The Davis-Putnam procedure~\cite{DavisPutnam60} is a well-known
algorithm that solves SAT. 
In its most basic form, it takes an ordering of the variables
$x_1,\ldots,x_n$ of a formula $F$ and checks whether
$\DP_{x_1,\ldots,x_n}(F)$ is empty or contains the empty clause.  In the
first case $F$ is satisfiable, and in the second case $F$ is
unsatisfiable.
Note that this procedure computes a certificate for the
(un)satisfiability of $F$; we can obtain a satisfying
truth assignment of $F$ from a satisfying truth assignment of $\DP_x(F)$, and 
we can obtain a resolution refutation of~$F$ from a resolution refutation of 
$\DP_x(F)$.
However, $\DP_x(F)$ contains in general more clauses than $F$. Hence, repeated
application of Davis-Putnam resolution to $F$ may cause an exponential
growth in the number of clauses. As a result, the Davis-Putnam procedure
has an exponential worst-case running time.

\subsection{Graphs and Hypergraphs}\label{s-graphs}

A \emph{hypergraph} $H$ is a pair $(V,E)$ where $V$ is the set of {\it
  vertices} and $E$ is the set of \emph{hyperedges}, which are subsets
of $V$. 
If $|e|=2$ then we call $e$ an \emph{edge}; we denote an edge
$e=\{u,v\}$ simply as $uv$ or $vu$.
If all hyperedges of a hypergraph are edges then we call it a 
\emph{graph}.  We say that a hypergraph $H'=(V',E')$ is a \emph{partial
  hypergraph} of $H=(V,E)$ if $V' \subseteq V$ and $E' \subseteq E$.
The \emph{incidence graph} $I(H)$ of hypergraph $H=(V,E)$ is the
bipartite graph with partition classes $V$ and $E$ such that  $e\in E$ is adjacent to $v\in V$ if and only if $v\in e$.  A
hypergraph is \emph{$\alpha$-acyclic} if it can be reduced to the empty
hypergraph by repeated application of the following rules:

\begin{enumerate}
\item Remove hyperedges that are empty or contained in other hyperedges.
\item Remove vertices that appear in at most one hyperedge.
\end{enumerate}

\noindent A hypergraph $H$ is \emph{$\beta$-acyclic} if every partial
hypergraph of $H$ is $\alpha$-acyclic. The \emph{hypergraph} $H(F)$ of
a formula
$F$ has vertex set $\var(F)$ and hyperedge set $\SB\var(C) \SM
C \in F\SE$. We say that $F$ is \emph{$\alpha$-acyclic} or {\it
  $\beta$-acyclic} if $H(F)$ is $\alpha$-acyclic or $\beta$-acyclic,
respectively. 

Let $F$ be a formula.
The \emph{incidence graph} of $F$ is the bipartite graph $I(F)$
with vertex set $\var(F)\cup F$ and edge set $\SB Cx \SM C\in F$ and
$x\in \var(C)\SE$.  
The \emph{directed incidence graph} of $F$ is the
directed graph $D(F)$ with vertex set $\var(F) \cup F$ and arc set $\SB (C,x)
\SM C \in F$ and $x \in C\SE \cup \SB (x,C) \SM C \in F$ and $\ol{x} \in
C\SE$. We can also represent the orientation of edges by labeling them
with the signs $+,-$, such that an edge between a variable $x$ and a
clause $C$ is labeled $+$ if $x\in C$ and labeled $-$ if $\ol{x}\in
C$. This gives rise to the \emph{signed incidence graph} which carries
exactly the same information as the directed incidence graph.

The graph parameter \emph{clique-width} measures in a certain
sense the structural complexity of a directed or undirected graph 
\cite{CourcelleEngelfrietRozenberg90}.  The parameter is defined via a
graph construction process where only a limited number of vertex labels
are available; vertices that share the same label at a certain point of
the construction process must be treated uniformly in subsequent
steps. In particular, one can use the following four operations: the
creation of a new vertex with label $i$, the vertex-disjoint union of
already constructed labeled graphs, the relabeling of all vertices of
label $i$ with label $j$ denoted $\rho_{i\rightarrow j}$, and the insertion of all possible edges
between vertices of label $i$ and label $j$ denoted $\eta_{i,j}$ (either undirected, in which case we can also write $\eta_{j,i}$,  
or directed from label $i$ to $j$). The clique-width $\cwd(G)$ of a graph
$G$ is the smallest number $k$ of labels that suffice to construct $G$
by means of these four operations.  
An algebraic term representing such a construction of $G$ 
is called a {\it $k$-expression} of $G$.
The \emph{(directed) clique-width} of a CNF formula is the clique-width
of its (directed) incidence graph.  The directed clique-width of a CNF
formula can also be defined in terms of the signed incidence graph and
is therefore sometimes called the \emph{signed clique-width}. 

Let $G=(V,E)$ be a graph. For a subset $U\subseteq V$, the subgraph of $G$ {\it induced by} $U$ is the graph with vertex set $U$ and edge set $\{uv\; |\; u,v\in U\; \mbox{with}\; uv\in E\}$.
A {\it cycle} is a graph, the vertices of which can be ordered as 
$v_1,\ldots,v_n$ such that $E=\{v_iv_{i+1}\; |\; 1\leq i\leq n-1\}\cup \{v_nv_1\}$.
A graph is {\it chordal bipartite} if it has no induced cycle on 6 vertices or more.  A vertex $v$ in a graph $G$ is \emph{weakly simplicial} if (i)~the
neighborhood of $v$ in $G$ forms an independent set, and (ii)~the
neighborhoods of the neighbors of $v$ form a chain under set
inclusion. 
Uehara~\cite{Uehara02} showed the following, which also
follows from results of Hammer, Maffray, and Preismann
\cite{HammerMaffrayPreismann89}, see \cite{PelsmajerTokazyWest04}.  We
call a bipartite graph \emph{nontrivial} if it contains at least one
  edge.

\begin{proposition}[\cite{HammerMaffrayPreismann89,Uehara02}]\label{prop:weakly-simplicial}
  A graph is chordal bipartite if and only if every induced subgraph has
  a weakly simplicial vertex. Moreover, a nontrivial chordal
  bipartite graph has a weakly simplicial vertex in each partition class.
\end{proposition}

The following proposition shows how $\beta$-acyclic
CNF formulas and chordal bipartite graphs are related.
The equivalence between statement~(i)
and~(ii) is due to Tarjan and
Yannakakis~\cite{TarjanYannakakis84}, 
who presented this relationship 
in terms of $\beta$-acyclic hypergraphs. 
The equivalence between
statement~(ii) and~(iii)  follows from the
facts that $I(H(F))$ is obtained from $I(F)$ after removing all but one
clause vertices in $I(F)$ with the same neighbors, i.e., clauses with
the same set of variables in $F$, and that a chordal bipartite graph
remains chordal bipartite under vertex deletion.

\begin{proposition}[\cite{TarjanYannakakis84}]\label{p-three}
For a CNF formula $F$, statements (i)-(iii) are equivalent:  
\begin{quote}
\begin{enumerate} 
\item [\normalfont (i)] 
$F$ is $\beta$-acyclic;
\item [\normalfont (ii)] 
$I(H(F))$ is chordal bipartite;
\item [\normalfont (iii)] 
$I(F)$ is chordal bipartite.
\end{enumerate}
\end{quote}
\end{proposition}

\noindent We also call a vertex of a hypergraph or a variable of a CNF formula 
{\it weakly simplicial} if the corresponding vertex in the associated incidence graph is weakly simplicial.

\section{Polynomial-time SAT Decision for $\beta$-acyclic CNF
  Formulas}\label{sec:poly}

Note that we can make a hypergraph $\alpha$-acyclic by adding a
universal hyperedge that contains all vertices; by rule 1 we remove all other hyperedges, by rule 2 all
vertices. By this observation, it is easy to see that SAT is \NP-complete
for the class of $\alpha$-acyclic CNF formulas~\cite{SamerSzeider10}. In
contrast, it is well known that the satisfiability of $\alpha$-acyclic
instances of the \textsc{Constraint Satisfaction Problem} (CSP) can be
decided in polynomial time~\cite{GottlobLeoneScarcello01a}.  Thus SAT
and CSP behave differently with respect to $\alpha$-acyclicity (representing a
clause with $k$ literals as a relational constraint requires exponential
space of order $k2^k$).
However, in this section, we give a polynomial-time algorithm that solves SAT 
for $\beta$-acyclic CNF formulas. 

If we can reduce a hypergraph $H$ to the empty graph by repeated deletion of 
weakly simplicial vertices, then we say that $H$ admits a \emph{weakly simplicial elimination ordering}. If $H=H(F)$ for some formula $F$, then we also say
that $F$ admits a \emph{weakly simplicial ordering} of its variables.
The first key ingredient of our algorithm is the following lemma.

\begin{lemma}\label{l-weakly}
If $F$ is a $\beta$-acyclic formula, then $F$ 
admits a weakly simplicial elimination ordering.
Moreover, such an ordering can be found in polynomial time.
\end{lemma}

\begin{proof}
Let $F$ be a $\beta$-acyclic formula. 
We must show that $H(F)$ admits a weakly simplicial elimination ordering.
Proposition~\ref{p-three} tells us that $I(H(F))$ is chordal bipartite.
Then $I(H(F))$ has a weakly simplicial vertex in each partition class due to
Proposition~\ref{prop:weakly-simplicial}. We choose the partition class of
$I(H(F))$ that corresponds to the vertices of $H$. 
Then the lemma readily follows after observing that the class of
chordal bipartite graphs is closed under vertex deletion and that
weakly simplicial vertices can be identified in polynomial time by
brute force.
\end{proof}

The following lemma is the second key ingredient for our
algorithm. Recall that $\DP_x(F)$ denotes the formula obtained from a formula $F$ after eliminating $x$ by Davis-Putnam resolution.

\begin{lemma}\label{lem:stays-small}
  If $x$ is a  weakly simplicial variable of a formula $F$, then
   $|\DP_x(F)| \leq |F|$. 
\end{lemma}
\begin{proof}
  Let $x$ be a weakly simplicial variable of a CNF formula $F$.
  Let $F-x:=\SB C\setminus \{x,\ol{x}\}\SM C\in F\SE $.
  We show that $\DP_x(F) \subseteq F-x$.

  Assume $C_1,C_2\in F$ have a resolvent $C$ with respect to $x$.
  Consequently we have $C_1\cap \ol{C_2} \subseteq \{x,\ol{x}\}$.
  Because $x$ is weakly simplicial, $\var(C_1)\subseteq \var(C_2)$ or
  $\var(C_2)\subseteq \var(C_1)$. Without loss of generality, assume the
  former is the case.  If $x\in C_1$, then we have $C_1\cap
  \ol{C_2}=\{x\}$, and so $C=C_2\setminus \{\ol{x}\}\in F-x$.
  Similarly, if $\ol{x}\in C_1$, then we have $C_1\cap
  \ol{C_2}=\{\ol{x}\}$, and so $C=C_2\setminus \{x\}\in F-x$.  Thus
  indeed $\DP_x(F) \subseteq F-x$.  From $\Card{\DP_x(F)}\leq
  \Card{F-x}\leq \Card{F}$ the result now follows.
\end{proof}

We are now ready to present our algorithm. 

\medskip
\noindent \rule{\textwidth}{0.1mm}\\
{\bf Algorithm solving} SAT {\bf for $\beta$-acyclic formulas}

\medskip
\noindent
\begin{tabular}[h]{lcl}
Input &:& a $\beta$-acyclic formula $F$\\[2pt]
Output &:& {\tt Yes} if $F$ is satisfiable\\
& & {\tt No} otherwise
\end{tabular}

\medskip
\noindent
{\bf Step 1.} compute a weakly simplicial elimination ordering $x_1,\ldots,x_n$ of $F$

\smallskip
\noindent
{\bf Step 2.} apply the Davis-Putnam procedure on ordering $x_1,\ldots,x_n$\\
\smallskip
\noindent \rule{\textwidth}{0.1mm}

\smallskip
We let $\BAC$ denote the class of all $\beta$-acyclic
formulas and state the main result of this section.

\begin{theorem}\label{the:beta-poly}
{\sc SAT} can be solved in polynomial time for $\BAC$.
\end{theorem}

\begin{proof}  
Let $F$ be a $\beta$-acyclic CNF formula.
We apply our algorithm. 
Its correctness follows from Lemma~\ref{l-weakly}
combined with
the correctness of the Davis-Putnam procedure~\cite{DavisPutnam60}. 
Steps 1 and 2 run in polynomial time due to Lemmas~\ref{l-weakly} and~\ref{lem:stays-small}, respectively. 
Hence, Theorem~\ref{the:beta-poly} follows.
\end{proof}

\section{Generalizing $\beta$-Acyclic Formulas}\label{s-gen}

Lemma~\ref{lem:stays-small} is one of the two key ingredients than ensures 
that our algorithm for solving SAT on $\BAC$ runs in polynomial time.
It states that the number of clauses does
not increase after applying Davis-Putnam resolution if $x$ is a weakly 
simplicial variable of a formula $F$. 
We can ensure this by requiring the following property
that is more general than being weakly simplicial. 
We say that a variable $x\in \var(F)$ is {\it DP-simplicial} in
a formula $F$ if

\begin{enumerate}
\item[(*)] for any two clauses $C,D\in
F$ that have an $x$-resolvent, this $x$-resolvent is a subset of $C$ or
a subset of $D$. 
\end{enumerate}

Observe that whenever an $x$-resolvent is a subset of a parent clause
$C$ then it is equal to $C\setminus \{x,\ol{x}\}$.
The following lemma immediately
follows from $(*)$.

\begin{lemma}\label{l-small2}
If $x$ is a  DP-simplicial variable of a formula $F$, then
$|\DP_x(F)| \leq |F|$.
\end{lemma}

An ordering $x_1,\dots,x_n$ of the variables of $F$ is a
\emph{DP-simplicial elimination ordering} if $x_i$ is DP-simplicial in
$\DP_{x_1,\dots,x_{i-1}}(F)$ for all $1\leq i \leq n$. 
We let $\DPS$
denote the class of all formulas that admit a DP-simplicial elimination
ordering.  We observe that every weakly simplicial elimination ordering
of $H(F)$ is a DP-simplicial elimination ordering of $F$.  This means
that $\BAC\subseteq \DPS$.  
However, due to Example~\ref{exa} below, the
reverse is not true.  Hence, we found the following result.

\begin{proposition}\label{p-proper0}
$\BAC \subsetneq \DPS$. 
\end{proposition}

Given a DP-simplicial ordering, the Davis-Putnam procedure runs in polynomial time due to Lemma~\ref{l-small2}. This leads to the following result.

\begin{proposition}\label{pro:DP}
{\sc SAT} can be solved in polynomial time for $\DPS$ provided that a DP-simplicial elimination ordering is given.
\end{proposition} 

\subsection{An Example}\label{exa}

We give an example of a formula in $\DPS\setminus \BAC$.  Consider the
formula $F$ that has variables $y$, $z$, $b$, $b'$, $b^*$ and $c$
and clauses $\{y,b,b^*,c\}$, $\{y,\ol{b}\}$, $\{y,\ol{b},b',z\}$,
$\{\ol{y},b,b',\ol{c}\}$, $\{\ol{y},\ol{b}\}$,
$\{\ol{y},\ol{b},b^*,\ol{z}\}$, $\{\ol{b},b'\}$, $\{\ol{b},z\}$,
$\{\ol{b},\ol{z}\}$, $\{b',b^*,c\}$, $\{b',b^*,\ol{c}\}$, $\{b',\ol{b^*}\}$, and
$\{\ol{b'},b^*\}$; see Figure~\ref{fig:example-formula} for an illustration.

\begin{figure}
\begin{tabular}{p{4.5cm}p{3.5cm}p{3cm}p{2.5cm}}

\begin{center}
  \begin{tabular}{cccccc}
    $y$ & $b$ & $b'$ & $b^*$ & $c$ & $z$ \\
    \hline
    $+$ & $+$  &0 & $+$ & $+$ & 0\\
    $+$ & $-$ & 0 &0 &0 &0 \\
    $+$ & $-$ & $+$ & 0 & 0 & $+$\\
    $-$ & $+$ & $+$ & 0& $-$ &0 \\
    $-$ & $-$ & 0&0 &0 &0 \\
    $-$ & $-$ & 0& $+$ &0 & $-$\\
    0& $-$ & $+$ &0 &0 &0 \\
    0& $-$ &0 &0 &0 & $+$ \\
    0& $-$ &0&0 &0 & $-$\\
    0&0 & $+$ & $+$ & $+$ &0 \\
    0&0 & $+$ & $+$ & $-$ &0\\
    0&0 & $+$ & $-$ &0 &0 \\
    0&0 & $-$ & $+$ &0 &0 \\
  \end{tabular}
\end{center}
&
\begin{center}
  \begin{tabular}{ccccc}
    $b$ & $b'$ & $b^*$ & $c$ & $z$ \\
    \hline
    $-$& $+$ &0 &0 & $+$\\
    $-$ &0 &0 &0 &0 \\
    $-$ &0 & $+$ &0 & $-$\\
    $-$ & $+$ &0 &0 &0 \\
    $-$ &0 &0 &0 & $+$ \\
    $-$ &0 &0 &0 & $-$ \\
    0& $+$ & $+$ & $+$ &0 \\
    0& $+$ & $+$ & $-$ &0 \\
    0& $+$ & $-$ &0 &0 \\
    0& $-$ & $+$ &0 &0 \\
  \end{tabular}
\end{center}
&
\begin{center}
  \begin{tabular}{cccc}
    $b'$ & $b^*$ & $c$ & $z$ \\
    \hline
   $+$ & $+$ & $+$ &0 \\
    $+$ & $+$ & $-$ &0 \\
    $+$ & $-$ &0 &0 \\
    $-$ & $+$ &0 &0 \\
  \end{tabular}
\end{center}
&
\begin{center}
  \begin{tabular}{ccc}
    $b^*$ & $c$ & $z$ \\
    \hline
    $+$& $+$ &0 \\
    $+$ & $-$ &0 \\
  \end{tabular}
\end{center}
\\
\\
\centering $F$
& 
\centering $\DP_y(F)$
&
\centering $\DP_{y,b}(F)$
&
\centering $\DP_{y,b,b'}(F)$
\end{tabular}
\caption{The example formula $F$ and the first three subformulas
  obtained from $F$ using the
  DP-simplicial elimination ordering $y,b,b',b^*,c,z$. The formulas
  are given as matrices in which each row corresponds to a clause of the
  formula and each column corresponds to a variable. 
  Each matrix entry is either ``$+$'', ``$-$'' or ``$0$''
  indicating whether a variable appears positively, negatively, or is absent,  respectively,
  in a clause.}
\label{fig:example-formula}
\end{figure}

We observe first that none of the variables of $F$ are weakly simplicial.
Consequently, there is no weakly simplicial elimination ordering
of~$F$. Hence $F\notin \BAC$. However, we will show below that $y$, $b$, $b'$, $b^*$, $c$, $z$ is a DP-simplicial elimination ordering of~$F$. Then
$F\in \DPS$, as desired (see Figure~\ref{fig:example-formula} for an illustration).

We find that $y$ is DP-simplicial in $F$ and obtain $ \DP_{y}(F) = \{
\{\ol{b},b',z\}$, $\{\ol{b}\}$, $\{\ol{b},b^*,\ol{z}\}$, $\{\ol{b},b'\}$,
$\{\ol{b},z\}$, $\{\ol{b},\ol{z}\}$, $\{b',b^*,c\}$,
$\{b',b^*,\ol{c}\}$, $\{b',\ol{b^*}\}$, $\{\ol{b'},b^*\} \}$.
We then find that $b$ is DP-simplicial in $\DP_y(F)$ and obtain $
\DP_{y,b}(F) = \{ \{b',b^*,c\}$, $\{b',b^*,\ol{c}\}$,
$\{b',\ol{b^*}\}$, $\{\ol{b'},b^*\} \}$.
We then find that $b'$ is DP-simplicial in
$\DP_{y,b}(F)$ and obtain
$\DP_{y,b,b'}(F) = \{\{b^*,c\}$, $\{b^*,\ol{c}\}\}$.
We then find that $b^*$ is DP-simplicial in
$\DP_{y,b,b'}(F)$ and obtain $
\DP_{y,b,b',b^*}(F) = \emptyset$.
Hence, $y$, $b$, $b'$, $b^*$, $c$, $z$ is a DP-simplicial elimination ordering of~$F$.

We note that $z$ is also 
DP-simplicial in $F$.
Suppose that we started with $z$ instead of $y$.
We first derive that $ \DP_{z}(F) = \{ \{y,b,b^*,c\}$, $\{y,\ol{b}\}$,
$\{y,\ol{b},b'\}$, $\{\ol{y},b,b',\ol{c}\}$, $\{\ol{y},\ol{b}\}$,
$\{\ol{y},\ol{b},b^*\}$, $\{\ol{b},b'\}$,
$\{b',b^*,c\}$, $\{b',b^*,\ol{c}\}$, $\{b',\ol{b^*}\}$,
$\{\ol{b'},b^*\} \}$.  In contrast to $\DP_{y}(F)$, the
clauses $\{y,b,b^*,c\}$ and $\{\ol{y},b,b',\ol{c}\}$ are still contained
in $\DP_{z}(F)$. This implies that $\DP_z(F)$ has no DP-simplicial
variables. Consequently, $F$ has no DP-simplicial elimination ordering
that starts with $z$.

We conclude that 
in contrast to weakly simplicial elimination orderings
it is important to choose the right variable when we
want to obtain a DP-simplicial elimination ordering. In the next section
we will extend 
this consideration
and show that making the right
choice is in fact an \NP-hard problem.

\section{Recognizing Formulas in ${\mathbf \DPS}$}\label{s-npcomplete}

We prove that the problem of testing whether a given CNF formula belongs to the class $\DPS$, i.e., admits a DP-simplicial elimination ordering, is \NP-complete.
This problem is in \NP, because we can check in polynomial time whether
an ordering of the variables of a CNF formula is a DP-simplicial elimination ordering.
In order to show \NP-hardness we reduce from SAT. 
In Section~\ref{s-gadget} we construct a CNF formula $F'$ from a given 
CNF formula $F$. We also show a number of properties of $F'$.
In Section~\ref{s-reduction} we use these properties to prove that $F$ 
is satisfiable if and only if $F'$ admits a DP-simplicial elimination ordering.

\subsection{The Gadget and its Properties}\label{s-gadget}

For a given CNF formula $F$ with variables $x_1,\ldots, x_n$ called the
{\it $x$-variables} and clauses $C_1,\ldots,C_m$, we construct a CNF
formula $F'$ as follows.  For every $x_i$ we introduce two variables
$y_i$ and $z_i$. We call these variables the $y$-variables and
$z$-variables, respectively.  For every $C_j$ we introduce a variable
$c_j$. We call these variables the $c$-variables. We also add three new
variables $b, b'$ and $b^*$ called the $b$-variables.  We let
$\mbox{var}(F')$ consist of all $b$-variables, $c$-variables,
$y$-variables, and $z$-variables.

Let $C_j$ be a clause of $F$. We replace every $x$-variable in $C$ by its associated $y$-variable if the occurrence of $x$ in $C$ is positive; otherwise we replace it by its associated $z$-variable. 
This yields a clause $D_j$. For instance, if $C_j=\{x_1,\ol{x_2},x_3\}$ then
$D_j=\{y_1,z_2,y_3\}$.

We let $F'$ consist of the following $6n+4m+3$ clauses:
\begin{itemize}
\item [$\bullet$] $\{y_i,\ol{b}\}$ and $\{\ol{y_i},\ol{b}\}$ for $i=1,\ldots,n$
called $by$-clauses \\[-6pt]
\item [$\bullet$] $\{z_i,\ol{b}\}$ and $\{\ol{z_i},\ol{b}\}$ for $i=1,\ldots,n$
called $bz$-clauses \\[-6pt]
\item [$\bullet$] $\{y_i,z_i,\ol{b},b'\}$ and $\{\ol{y_i},\ol{z_i},\ol{b},b^*\}$ for $i=1,\ldots,n$ called $byz$-clauses\\[-6pt]
\item [$\bullet$] $\{c_j,b',b^*\}$ and $\{\ol{c_j},b',b^*\}$ for $j=1,\ldots,m$
called $bc$-clauses\\[-6pt]
\item [$\bullet$] $D_j\cup \{b,b^*,c_j\}\cup \SB\ol{c_k} \SM k\neq j\SE$
  and $\ol{D_j}\cup \{b,b',c_j\}\cup \SB\ol{c_k} \SM k\neq j\SE$ for $j=1,\ldots,m$\\[3pt] called $bcD$-clauses\\[-6pt]
\item [$\bullet$] $\{\ol{b},b'\}$, $\{\ol{b'},b^*\}$ 
and $\{b',\ol{b^*}\}$ called $b$-clauses.
\end{itemize}

We call a pair $D_j\cup \{b,b^*,c_j\}\cup \SB\ol{c_k} \SM k\neq j\SE$
and $\ol{D_j}\cup \{b,b',c_j\}\cup \SB \ol{c_k} \SM k\neq j\SE$ for some $1\leq j\leq m$ a {\it $bcD$-clause pair.}
We call a CNF formula $M$ a \emph{$yz$-reduction formula} of $F'$ if
there exists a sequence of variables $v^1,\ldots,v^k$, where every $v^i$ is either a $y$-variable or a $z$-variable, such that
$\DP_{v^1,\dots,v^k}(F')$ $=$ $M$, and $v^i$ is DP-simplicial in
$\DP_{v^1,\dots,v^{i-1}}(F')$ for $i=1,\ldots,k$.
We say that two clauses $C$ and $D$ {\it violate} (*) if
they have a resolvent that is neither a subset of $C$ nor a subset of $D$, i.e.,
$C\cap \ol{D}=\{v\}$ for some variable $v$ but 
neither $(C\cup D)\setminus \{v,\ol{v}\}=C\setminus \{v\}$ nor $(C\cup
D)\setminus \{v,\ol{v}\}= D\setminus \{\ol{v}\}$.  We will now prove
five useful lemmas valid for $yz$-reduction formulas.

\begin{lemma}\label{l-dp-simplicial}
  Let $M$ be a $yz$-reduction formula of $F'$. If $M$ contains both clauses of some $bcD$-clause pair, then neither any $b$-variable nor any $c$-variable
  is DP-simplicial in $M$.
\end{lemma}
\begin{proof}
Let $E_1=D_j\cup \{b,b^*,c_j\}\cup \SB\ol{c_k} \SM k\neq j\SE$
and $E_2=\ol{D_j}\cup \{b,b',c_j\}\cup \SB\ol{c_k} \SM k\neq j\SE$ for some $1\leq j\leq m$ be a $bcD$-clause pair in $M$.
We observe that by definition $M$ contains all $b$-clauses and $bc$-clauses. 
This enables us to prove the lemma.
Let $v$ be a $b$-variable or $c$-variable. Then we must distinguish 5 cases.
If $v=b$, then $\{\ol{b},b'\}$ and $E_1$ violate (*). 
If $v=b'$, then $\{\ol{b'},b^*\}$ and $E_2$ violate (*).
If $v=b^*$, then $\{b',\ol{b^*}\}$ and $E_1$ violate (*).
If $v=c_j$, then $\{\ol{c_j},b',b^*\}$ and $E_1$ violate (*).
If $v=c_k$ for some $1\leq k\leq m$ with $k\neq j$, then $\{c_k,b',b^*\}$ and  $E_1$ violate (*).
\end{proof}

\begin{lemma}\label{l-at-most-one} 
  Let $M$ be a $yz$-reduction formula of $F'$.  
  Then $y_i \in \var(M)$ or $z_i\in \var(M)$ for $i=1,\ldots,n$.
\end{lemma}
\begin{proof}
Suppose that $M$ does not contain $y_i$ or $z_i$ for some $1\leq i\leq m$, say
$y_i\notin \var(M)$. We show that $z_i\in \var(M)$.
Let $M'$ be the formula obtained from $F'$ just before the elimination of
$y_i$. Because $M$ is a $yz$-reduction formula, $M'$ is a $yz$-reduction formula as well. Hence, $\var(M')$ contains all $b$-variables. Because $y_i$ and $z_i$ are in $\var(M')$, 
we then find that $M'$ contains the clauses 
$\{y_i,z_i,\ol{b},b'\}$, $\{\ol{y_i},\ol{b}\}$, $\{\ol{y_i},\ol{z_i},\ol{b},b^*\}$ and $\{y_i,\ol{b}\}$. Because the first two clauses resolve into $\{z_i,\ol{b},b'\}$, and the last two resolve into $\{\ol{z_i},\ol{b},b^*\}$, 
we obtain that $\DP_{y_i}(M')$ contains 
$\{z_i,\ol{b},b'\}$ and $\{\ol{z_i},\ol{b},b^*\}$, which violate (*).
Because $M$ contains all $b$-variables by definition, $z_i$ will never become
DP-simplicial when we process $\DP_{y_i}(M')$ until we obtain $M$. Hence, $z_i\in \var(M)$, as desired.
\end{proof}

\begin{lemma}\label{l-clauses-sat} 
  Let $M$ be a $yz$-reduction formula of $F'$, and let $1\leq j\leq m$.
  If there is a variable that occurs in $D_j$ but not in $M$, then
  $M$ neither contains 
  $D_j\cup \{b,b^*,c_j\}\cup \SB\ol{c_k}\SM k\neq j\SE$
  nor $\ol{D_j}\cup \{b,b',c_j\}\cup \SB\ol{c_k} \SM k\neq j\SE$ nor their  resolution descendants.
\end{lemma}
\begin{proof}
Let $v$ be a variable that occurs in $D_j$ but not in $M$. We may assume without loss of generality that $v$ is the first variable in $D_j$ that got eliminated and that $v=y_i$ for some $1\leq i\leq n$.
Let $S$ be the set that consists of all clauses $D_{j'}\cup
\{b,b^*,c_{j'}\}\cup 
\SB \ol{c_k}\SM k\neq j'\SE$ and $\ol{D_{j'}}\cup \{b,b',c_{j'}\}\cup
\SB\ol{c_k} \SM k\neq j'\SE$ in which $y_i$ occurs. 

Let $M'$ be the formula obtained from $F'$ just before the elimination of $y_i$.
Because $M$ is a $yz$-reduction formula, $M'$ is a $yz$-reduction formula as well. Hence, by definition, all $b$-variables and all $c$-variables occur in~$M'$. Then the clauses in $M'$, in which $y_i$ occurs, are
$\{y_i,\ol{b}\}$, $\{\ol{y_i},\ol{b}\}$, $\{y_i,z_i,\ol{b},b'\}$,$\{\ol{y_i},\ol{z_i},\ol{b},b^*\}$, together with 
clauses that are either from $S$ or resolution descendants of clauses in $S$.
Note that these resolution descendants still contain all their
$b$-variables and $c$-variables.

When we eliminate $y_i$, we remove all clauses in $M'$ in which $y_i$ occurs.
Hence, $\DP_{y_i}(M')$, and consequently, $M$ neither contains 
$E_1=D_j\cup \{b,b^*,c_j\}\cup \SB \ol{c_k}\SM k\neq j\SE$
nor $E_2=\ol{D_j}\cup \{b,b',c_j\}\cup \SB \ol{c_k}\SM k\neq j\SE$.
We show that $\DP_{y_i}(M')$ does not contain a resolvent of one of these two clauses either. This means that $M'$ does not contain one of their resolution descendants, as desired.
We only consider $E_1$, because we can deal with $E_2$ in the same way. There is no $y_i$-resolvent of $E_1$ and a clause $C$ from 
$\{\{y_i,\ol{b}\}$, $\{\ol{y_i},\ol{b}\}$, $\{y_i,z_i,\ol{b},b'\}$,$\{\ol{y_i},\ol{z_i},\ol{b},b^*\}\}$, because $E_1\cap \ol{C}$ contains $b$.
There is no $y_i$-resolvent of $E_1$ and a (resolution descendant from a) clause 
$C$ of $S$ either, because $E_1\cap \ol{C}$ contains $c_j$.
\end{proof}

\begin{lemma}\label{l-variable-simplicial} 
  Let $M$ be a $yz$-reduction formula of $F'$, and let $1\leq i\leq n$. 
  If $\var(M)$ contains $y_i$ and $z_i$, then both $y_i$ and $z_i$ are DP-simplicial in $M$.
\end{lemma}
\begin{proof}
  By symmetry, we only have to show that $y_i$ is DP-simplicial in $M$.
  Let $S$ be the set of all clauses $D_{j'}\cup \{b,b^*,c_{j'}\}\cup
  \SB\ol{c_k}\SM k\neq j'\SE$ and $\ol{D_{j'}}\cup
  \{b,b',c_{j'}\}\cup \SB\ol{c_k}\SM k\neq j'\SE$ in which $y_i$
  occurs.  By definition, $\var(M)$ contains all $b$-variables and all
  $c$-variables.  This has the following two consequences.  First, as
  $\var(M)$ also contains $y_i$ and $z_i$, we find that $M$ contains the
  clauses $\{y_i,\ol{b}\}$, $\{\ol{y_i},\ol{b}\}$,
  $\{y_i,z_i,\ol{b},b'\}$, and $\{\ol{y_i},\ol{z_i},\ol{b},b^*\}$.
  Second, by Lemma~\ref{l-clauses-sat}, the other clauses of $M$ in
  which $y_i$ occurs form a subset of $S$.  This means that there are
  only 3 pairs of clauses $C_1,C_2$ in $M$ with $C_1\cap
  \ol{C_2}=\{y_i\}$, namely the pair $\{y_i,\ol{b}\}$,
  $\{\ol{y_i},\ol{b}\}$, the pair $\{y_i,\ol{b}\}$,
  $\{\ol{y_i},\ol{z_i},\ol{b},b^*\}$, and the pair
  $\{\ol{y_i},\ol{b}\}$, $\{y_i,z_i,\ol{b},b'\}$.  Each of these pairs
  satisfies (*). This completes the proof of
  Lemma~\ref{l-variable-simplicial}.
\end{proof}

\begin{lemma}\label{l-end-sequence} 
  Let $M$ be a $yz$-reduction formula of $F'$.
  If $M$ contains neither $bcD$-clauses nor resolution descendants of such
  clauses, then $M$ has a DP-simplicial elimination ordering
  $b,c_1,\dots,c_m,b',b^*,v^1,\dots,v^\ell$, where $v^1,\ldots,v^\ell$ form an arbitrary ordering of the $y$-variables and $z$-variables in $\var(M)$.
\end{lemma}
\begin{proof}
By our assumptions, the only clauses in $M$ in which $b$ occurs are
$by$-clauses, $bz$-clauses, $byz$-clauses, and the clause $\{\ol{b},b'\}$. 
In all these clauses $b$ occurs as $\ol{b}$.
Hence, $b$ is (trivially) DP-simplicial in $M$. 
We then find that $\DP_b(M)$ consists of 
$\{\ol{b'},b^*\}$, $\{b',\ol{b^*}\}$ and all $bc$-clauses.
For every $c_j$, there exists exactly one $bc$-clause, namely
$\{c_j,b',b^*\}$, in which $c_j$ occurs as $c_j$, and exactly one $bc$-clause, namely $\{\ol{c_j},b',b^*\}$, in which $c_j$ occurs as $\ol{c_j}$.
Hence, 
$c_j$ is DP-simplicial in
$\DP_{b,c_1,\dots,c_{j-1}}(M)$ for $j=1,\ldots, m$.
We deduce that $\DP_{b,c^1,\dots,c^m}(M)=\{\{b',b^*\}, \{\ol{b'},b^*\},
\{b',\ol{b^*}\}\}$. Then $b'$ is DP-simplicial in
$\DP_{b,c_1,\dots,c_m}(M)$, and we find that
$\DP_{b,c_1,\dots,c_m,b'}(M)=\{\{b^*\}\}$. Then $b^*$ is DP-simplicial
in $\DP_{b,c_1,\dots,c_m,b'}(M)$, and we find that $\DP_{b,c_1,\dots,c_m,b',b^*}(M)=\emptyset$. 
Consequently, 
$v^i$ is DP-simplicial in
  $\DP_{b,c_1,\dots,c_m,b',b^*,v^1,\dots,v^{i-1}}(M)$ for $i=1,\ldots,\ell$.  This concludes
  the proof of Lemma~\ref{l-end-sequence}.
\end{proof}

\subsection{The Reduction}\label{s-reduction}

We are now ready to prove the main result of Section~\ref{s-npcomplete}.

\begin{theorem}\label{the:hardness-dps}
  The problem of testing whether a CNF formula belongs to $\DPS$
  is \NP-complete.
\end{theorem}
\begin{proof}
  Recall that the problem is in \NP.  Given a CNF formula $F$
  that has variables $x_1,\ldots,x_n$ and clauses $C_1,\ldots,C_m$, we
  construct in polynomial time the CNF formula~$F'$.  We claim that $F$
  is satisfiable if and only if~$F'$ admits a DP-simplicial elimination
  ordering.

  First suppose that $F$ is satisfiable.  Let $\tau$ be a satisfying
  truth assignment of $F$.  We define functions $f$ and $g$ that map
  every $x$-variable to a $y$-variable or $z$-variable in the following
  way.  If $\tau(x_i)=1$, then $f(x_i)=y_i$ and $g(x_i)=z_i$. If
  $\tau(x_i)=0$, then $f(x_i)=z_i$ and $g(x_i)=y_i$.  Let
  $x_1,\dots,x_n$ be the $x$-variables in an arbitrary ordering.  Then,
  for every $1 \leq i \leq n$, the formula
  $\DP_{f(x_1),\dots,f(x_i)}(F')$ is a $yz$-reduction formula.  From
  Lemma~\ref{l-variable-simplicial} we deduce that $f(x_i)$ is
  DP-simplicial in $\DP_{ f(x_1),\dots,f(x_{i-1}) }(F')$ for every
  $1\leq i\leq n$.  Because $\tau$ satisfies $F$, $\var(D_j)$ contains a
  variable that is not in $\var(\DP_{f(x_1),\dots,f(x_n)}(F'))$, for
  every $1\leq j \leq m$.  Lemma~\ref{l-clauses-sat} implies that
  $M$ does not contain any $bcD$-clause or any of their resolution
  descendants.  Then, by Lemma~\ref{l-end-sequence}, we find that $
  f(x_1),\dots,f(x_n), b, c_1,\dots,c_m, b',b^*, g(x_1),\dots,g(x_n) $
  is a DP-simplicial elimination ordering of $F'$.

Now suppose that $F'$ admits a DP-simplicial elimination ordering
$v^1,\dots,v^{\Card{\var(F')}}$.
Let $v^k$ be the first variable that is neither a $y$-variable nor a $z$-variable.
Then $M=\DP_{v^1,\dots,v^{k-1}}(F')$ is a $yz$-reduction formula.
Let $A=\{v^1,\dots,v^{k-1}\}$, and let $X$ consist of all $x$-variables that have
an associated $y$-variable or $z$-variable in $A$. 
We define a truth assignment $\tau:X\rightarrow \{0,1\}$ by setting $\tau(x_i)=1$ if $y_i \in A$ and $\tau(x_i)=0$ if $z_i \in A$, for every $x_i\in X$.  By Lemma~\ref{l-at-most-one},
we find that $\tau$ is well defined.  Because $v^k$ is a DP-simplicial $b$-variable or a DP-simplicial $c$-variable in $M$, we can apply Lemma~\ref{l-dp-simplicial} and find that, for every $1\leq j\leq m$, at least one of the two
clauses 
$D_j\cup \{b,b^*,c_j\}\cup \SB\ol{c_k}\SM k\neq j\SE$
and $\ol{D_j}\cup \{b,b',c_j\}\cup \SB \ol{c_k}\SM k\neq j\SE$
is not in $M$. This means that every clause $C_j$ contains a literal $x$
with $\tau(x)=1$. Hence, $F$ is satisfiable. This completes the proof of Theorem~\ref{the:hardness-dps}. 
\end{proof}

\section{Intermediate Classes}\label{s-inter}

We discuss a possibility for coping with the 
\NP-hardness result of the previous section. 
The ultimate reason for this
hardness is that a formula may have several DP-simplicial variables, and
it is hard to choose the right one. A simple workaround is to assume a
fixed ordering of the variables and always choose the DP-simplicial
variable which comes first according to this ordering. 
In this way we loose some generality but win polynomial time tractability. This idea is made explicit in the following definitions.

Let $\Omega$ denote the set of all strict total orderings of the
propositional variables.  Let $\mathnormal{\prec}\in \Omega$ and $F$ 
be a CNF formula. A variable $x\in \var(F)$ 
is \emph{$\prec$-DP-simplicial} 
in $F$ if $x$ is DP-simplicial 
in $F$, and $\var(F)$ contains no
variable $y\prec x$ that is DP-simplicial in $F$. 
A strict total ordering $x_1,\dots,x_n$ of the variables of $F$ is a
\emph{$\prec$-DP-simplicial elimination ordering} if $x_i$ is
$\prec$-DP-simplicial in $\DP_{x_1,\dots,x_{i-1}}(F)$ for all $1\leq i \leq n$.
We let $\DPS_\prec$ denote the class of all CNF formulas that admit a $\prec$-DP-simplicial elimination ordering, and we set
$\DPS_\forall=\bigcap_{\mathnormal{\prec}\in \Omega} \DPS_\prec$.

\begin{proposition}
  $\DPS_\prec$ can be recognized in polynomial time for every
  $\mathnormal{\prec}\in \Omega$. 
  More precisely,
  it is possible to find in polynomial time a $\prec$-DP-simplicial
  elimination ordering for a given CNF formula $F$, or else to decide that
  $F$ has no such ordering.
\end{proposition}  
\begin{proof}
  Let $x_1,\dots,x_n$ be the variables of $F$, ordered according to
  $\prec$.    
  By brute force we check whether $x_i$ is DP-simplicial in $F$, for
  $i=1,\dots,n$. This takes polynomial time for each check.
  When we have found
  the first DP-simplicial variable $x_i$, we replace $F$   
  by
  $\DP_{x_i}(F)$. We iterate this procedure as long as possible.  Let
  $F'$ be the formula we end up with. If $\var(F')=\emptyset$ then $F\in
  \DPS_\prec$ and the sequence of variables as they have been eliminated
  provides a $\prec$-DP-simplicial elimination ordering.  If
  $\var(F')\neq\emptyset$ then $F\notin \DPS_\prec$.
\end{proof}

\begin{proposition}\label{pro:chain}
  $\BAC \subsetneq \DPS_\forall \subsetneq \DPS =
  \bigcup_{\mathnormal{\prec}\in \Omega} \DPS_\prec$.
\end{proposition}
\begin{proof}
  First we show that $\BAC \subsetneq \DPS_\forall$.  Let $F\in \BAC$
  and $\mathnormal{\prec}\in \Omega$.  We use induction on the
  number of variables of $F$ to show that $F\in \DPS_\prec$. 
The base case $|\var(F)|=0$ is trivial. 
Let $|\var(F)|\geq 1$.  
Because $F\in \BAC$ and $\var(F)\neq\emptyset$,
we find that $F$ has at least one weakly simplicial variable.  
Recall that each weakly simplicial variable is DP-simplicial. 
Consequently, $F$ has at least one DP-simplicial variable. 
Let $x$ be the first DP-simplicial variable in the ordering
$\prec$. By definition, $x$ is a $\prec$-DP-simplicial variable. We consider
  $F'=\DP_x(F)$. Because a $\beta$\hy acyclic hypergraph remains
  $\beta$-acyclic under vertex and hyperedge deletion, 
  $F'\in \BAC$. Because $F'$ has fewer variables than $F$, we 
  use the induction hypothesis to conclude that $F'\in \DPS_\prec$.  Hence
  $\BAC\subseteq \DPS_\prec$ follows.  
  Because $\mathnormal{\prec}\in
  \Omega$ was chosen arbitrarily, $\BAC \subseteq \DPS_\forall$ follows.
 
 In order to see that $\BAC \neq \DPS_\forall$, we take a hypergraph $H$ that is
  not $\beta$-acyclic and consider $H$ as a CNF formula with only positive
  clauses. All variables of $H$ are DP-simplicial and can be eliminated
  in an arbitrary order. Thus $H\in \DPS_\forall \setminus \BAC$.

  Next we show that $\DPS_\forall \subsetneq \DPS$. Inclusion holds by
  definition. In order to show that the inclusion is strict, we consider the
  formula $F$ of the example in Section~\ref{exa}.
  In that section we showed that $y$, $b$, $b'$, $b^*$, $c$, $z$ is a DP-simplicial elimination ordering of~$F$. Hence,
  $F\in \DPS_\prec$ for any ordering $\prec$ with $y\prec b\prec b' \prec b^* \prec c \prec z$.
  We also showed that $z$ is DP-simplicial in $F$ but that $F$ has no DP-simplicial ordering
   starting with $z$. Hence, $F\notin \DPS_{\prec'}$ for any ordering
  $\prec'$ with $z\prec' y$. We conclude that $F\in \DPS\setminus\DPS_\forall$.
Finally,  the equality
$\DPS = \bigcup_{\mathnormal{\prec}\in \Omega}
  \DPS_\prec$ holds by definition.
\end{proof}

\subsection{Grades of Tractability}

What properties do we require from a class $\CCC$ of CNF formulas to be
a ``tractable class'' for SAT? Clearly we want $\CCC$ to satisfy the property:
\begin{enumerate}
\item[1.] Given a formula $F\in \CCC$, we can decide in polynomial
  time whether $F$ is satisfiable.
\end{enumerate}
This  alone is not enough, since even the class of all
satisfiable CNF formulas has this property. 
Therefore we might wish that a tractable class 
$\CCC$ should also satisfy the property:
\begin{enumerate}
\item[2.] Given a formula $F$, we can decide in polynomial time
  whether $F\in \CCC$.
\end{enumerate}
However, if $\CCC$ is not known to satisfy
property~2, then it may still satisfy the property:
\begin{enumerate}
\item[3.] There exists a polynomial-time algorithm that either decides
whether a given a formula $F$ is satisfiable or not, 
or else decides that $F$ does not belong to~$\CCC$.
\end{enumerate}
The algorithm mentioned in property 3
may decide the satisfiability of some
formulas outside of $\CCC$, hereby avoiding the recognition problem. Such
algorithms are called \emph{robust algorithms}~\cite{Spinrad03}. 
In addition we would also assume from a tractable class $\CCC$ to be closed
under isomorphisms, i.e., to satisfy the property:
\begin{enumerate}
\item[4.] If two formulas differ only in the names of their variables, then
  either both or none belong to $\CCC$.
\end{enumerate}
This leaves us with two notions of a tractable class for SAT, a
\emph{strict} one where properties 1, 2, and 4 are required, and a
\emph{permissive} one where only properties 3 and 4 are required.  Every
strict class is permissive, but the converse does not hold in general (unless $\P=\NP$).
For instance, the class of Horn formulas is strictly tractable, but the
class of extended Horn formulas is only known to be
permissively tractable~\cite{SchlipfEtal95}.

Where are the classes from our paper located within this
classification?  
As a result of Theorem~\ref{the:beta-poly}, we find that
$\BAC$ is strictly tractable.
By Theorem~\ref{the:hardness-dps}, $\DPS$ is not strictly tractable (unless $\P
= \NP$).  The classes $\DPS_\prec$ do not satisfy property~4. Hence they are
not considered as tractable classes. However, $\DPS_\forall$ is
permissively tractable, because an algorithm for $\DPS_\prec$ for an
arbitrary ordering $\prec$ is a robust algorithm for $\DPS_\forall$. It
remains open whether $\DPS$ is permissively tractable.

\section{Comparisons}\label{s-comp}

We compare the classes of our paper with other known
(strictly or permissively) tractable classes.  
We say that two classes $\CCC_1$ and $\CCC_2$ of CNF
formulas are \emph{incomparable} if 
for every $n$ larger than some fixed constant there exist
formulas in $\CCC_1 \setminus \CCC_2$ and in $\CCC_2 \setminus \CCC_1$
with at least $n$ variables. 

We show that each of the classes mentioned in
Proposition~\ref{pro:chain} is incomparable with a wide range of
classes of CNF formulas, in particular with all the tractable classes
considered in~Speckenmeyer's survey~\cite{Speckenmeyer09}, and classes
based on graph width parameters~\cite{GottlobSzeider08}.  
For showing
this it suffices to consider the classes $\BAC$ and
$\DPS$ only, which are boundary classes as shown in Proposition~\ref{pro:chain}.

The following four families of formulas will be sufficient for
showing most of our incomparability results.
Here, $n \geq 1$ is an integer,
$x_1,\dotso,x_n$ and $y_1,\dotso,y_{2^n}$ are variables, and $C_1,
\dotso, C_{2^n}$ are all possible clauses with variables
$x_1,\ldots,x_n$. 

\begin{eqnarray*}
  \FALL(n) & = & \{C_1,\dotso,C_{2^n}\} \\[3pt]
  \FTWO(n) & = & \{\{x_1,\dotso,x_{\lceil \frac{n}{2} \rceil}\},\{x_{\lceil
    \frac{n}{2} \rceil},\dotso,x_n\}\} \\[3pt]
  \FCYC(n) & = & \SB \{x_i,\ol{x}_{i+1}\} \SM 1 \leq i \leq n-1 \SE \cup
  \{\{x_n,\ol{x}_1\}\} \\[3pt]
  \FALLCYC(n) & = & 
   \SB \{y_{j-1},\ol{y}_{j}\} \cup C_j \SM 1 < j \leq 2^n
  \SE \cup \{\{y_{2^n},\ol{y}_1\} \cup C_{1}\} \cup \\
  & & 
  \SB \{y_j,y_{j+1}\} \cup C_j \SM 1 \leq j \leq 2^n  \SE \cup
  \{\{y_{2^n},y_1\} \cup C_{2^n}\}.
\end{eqnarray*}

We observe that every $I(F_a(n))$ is  a complete bipartite graph with partition
classes of size $n$ and $2^n$, respectively, and 
that every $I(F_s(n))$ is a tree. Because complete bipartite graphs and trees  are chordal bipartite, we can apply Proposition~\ref{p-three} to obtain the following lemma. 

\begin{lemma}\label{lem:fall-ftwo}
  $\FALL(n),\FTWO(n) \in \BAC$ for all $n \geq 1$.
\end{lemma}

\noindent
By the following lemma, the other two classes of formulas do not intersect with $\DPS$. 
Recall that two clauses $C$ and $D$ violate (*) if
they have a resolvent that is neither a subset of $C$ nor a subset of $D$.

\begin{lemma}\label{lem:fcyc-fallcyc}
  $\FCYC(n),\FALLCYC(n) \notin \DPS$ for all $n \geq 3$.
\end{lemma}

\begin{proof}
Throughout the proof we compute indices of modulo $n$ for the
vertices $x_i$, and modulo $2^{n+1}$ for the vertices $y_j$.

First we show that $\FCYC(n) \notin \DPS$. 
The  clauses $C=\{x_i$, $\ol{x}_{i+1}\}$ and $C'=\{x_{i-1},\ol{x_i}\} \in \FCYC(n)$ 
have the $x_i$-resolvent $\{ x_{i-1}, \ol{x}_{i+1}\}$ which is not a subset of $C$ or $C'$.
  Hence, $C$ and $C'$ violate (*).  Consequently, $x_i$ is not $\DP$\hy simplicial
  for any $1\leq i \leq n$.
Because $\FCYC(n)$ has no other resolvents, $\FCYC(n)$ has no $\DP$-simplicial
variables. Because $\var(\FCYC(n))\neq \emptyset$ either,
we conclude that  $\FCYC(n) \notin \DPS$ for all $n\geq 3$.

Next we show that $\FALLCYC(n) \notin \DPS$.  
 Let $1\leq i \leq n$ for some $n\geq 3$. Let $1\leq j_1,j_2 \leq 2^n$ such that $C_{j_1} \cap
  \ol{C_{j_2}}=\{x_i\}$. By definition, $\FALLCYC(n)$ contains the 
  clauses $C=\{y_{j_1},y_{j_1+1}\} \cup C_{j_1}$ and
  $C'=\{y_{j_2},y_{j_2+1}\} \cup C_{j_2}$, which have 
  $x_i$-resolvent $C^*=\{y_{j_1},
  y_{j_1+1}, y_{j_2}, y_{j_2+1}\}\cup (C_{j_1}\cup C_{j_2})\setminus 
 \{x_i,\ol{x}_i\}$ .  However, since $\{y_{j_1}$, $y_{j_1+1}\}\neq \{y_{j_2}$, $y_{j_2+1}\}$, 
  we find that $C^*$ is not a subset of $C$ or $C'$. Hence,
 $C$ and $C'$ violate (*). 
Consequently,  $x_i$ is not $\DP$\hy simplicial for any $1\leq i \leq n$.

 Let $1 \leq j \leq 2^n$ for some $n\geq 3$. 
 Then $\FALLCYC(n)$ contains the two clauses $C=\{ y_j,
  y_{j+1}\} \cup C_j$ and $C'=\{y_{j-1},\ol{y}_j\} \cup C_j$, which 
  have $y_j$-resolvent $C^*=\{y_{j-1},y_{j+1}\}\cup C_j$.  However, $y_{j-1} \in C^*\setminus C$ and
  $y_{j+1}\in C^*\setminus C'$. Hence, $C^*$ is not a subset of
  $C$ or $C'$. Consequently $y_j$ is not $\DP$\hy simplicial for any $1
  \leq j \leq 2^n$. 
 Because $\FALLCYC(n)$ has no other resolvents, 
 $\FALLCYC(n)$ has no $\DP$-simplicial
variables. Because $\var(\FALLCYC(n))\neq \emptyset$ either,
 we conclude that  $\FALLCYC(n)
  \notin \DPS$ for all $n\geq 3$.
\end{proof}

\medskip
Suppose that we want to show that  $\BAC$ and $\DPS$ are incomparable with a class
$\CCC$ of CNF formulas. Then,
Proposition~\ref{pro:chain} combined with Lemmas~\ref{lem:fall-ftwo} and~\ref{lem:fcyc-fallcyc} implies that we only have to show the validity of  
the following two statements:
\begin{itemize}
\item [(i)] $\FALL(n)\notin \CCC$ or $\FTWO(n)\notin \CCC$ for every $n$ larger than some fixed constant;
\item [(ii)] $\FCYC(n)\in \CCC$ or $\FALLCYC(n)\in
\CCC$ for every $n$ larger than some fixed constant.
\end{itemize}

\subsection{Easy Classes}

We use (i) and (ii) to show that $\BAC$ and $\DPS$ are incomparable with the classes considered by Speckenmeyer~\cite{Speckenmeyer09}.  
For example, consider the class of \emph{2-CNF formulas}, i.e., CNF
formulas where every clause contains at most two literals. For
every $n \geq 3$, $\FALL(n)$ is not a 2-CNF formula. This
shows~(i). Furthermore, (ii) follows from the fact that $\FCYC(n)$ is a
2-CNF formula for every $n \geq 3$. Consequently, the class of 2-CNF
formulas is incomparable with $\BAC$ and $\DPS$.

As a second example we consider the class of \emph{hitting formulas},
i.e.,  CNF formulas where $C \cap \ol{C'} \neq \emptyset$ holds for
any two of their clauses \cite{Speckenmeyer09}. Now, for every $n \geq
3$ the formula $\FTWO(n)$ is not a hitting formula. This
shows~(i). It is not difficult to see that for $n \geq
3$, $\FALLCYC(n)$ is a hitting formula.  This shows~(ii). Consequently, the
class of hitting formulas is incomparable with $\BAC$ and $\DPS$.

The proofs for other classes of formulas considered
in~\cite{Speckenmeyer09} are similar. In particular, for the classes
\emph{Horn, renameable Horn, extended Horn, CC-balanced, Q-Horn, SLUR,
  Matched, bounded deficiency, nested, co-nested, and BRLR${}_k$
  formulas} we can utilize the formulas~$\FALL(n)$ to show~(i) and the
formulas~$\FCYC(n)$ to show~(ii).

\subsection{Classes of Bounded Width}
It is known~\cite{GottlobSzeider08} 
that SAT is tractable for various classes of formulas
that are defined by bounding certain width-measures of graphs associated
with formulas.
Besides the incidence graph $I(F)$ and the directed incidence graph $D(F)$, the 
other prominent graph associated with a CNF formula $F$ is
the \emph{primal graph} $P(F)$ of $F$, which is the graph with vertex set $\var(F)$ and
edge set $\SB x,y \SM x,y\in \var(C)\; \mbox{for some}\; C\SE$.
We restrict our scope to the graph invariants \emph{treewidth} (tw), and
clique-width (cw). Recall that the latter notion has been defined in Section~\ref{sec:pre}. 
For the definition of treewidth we refer to other
sources~\cite{GottlobSzeider08}, 
as we do not need it here.

For a graph invariant $\pi$, a graph representation $G\in
\{P,I,D\}$ and an integer~$k$, we consider the class $\CNF^G_k(\pi)$
of CNF formulas $F$ with $\pi(G(F))\leq k$. 
For every fixed $k\geq 0$,
SAT can be solved in polynomial time for the classes
$\CNF^P_k($tw$)$, $\CNF^I_k($tw$)$, and
$\CNF^D_k($cw$)$~\cite{GottlobSzeider08}. We show that these classes
are incomparable with $\BAC$ and $\DPS$.

\begin{proposition}\label{pro:tw}
  For every $k\geq 2$, $\CNF^P_k(\text{tw})$ is incomparable with $\BAC$
  and $\DPS$.
\end{proposition}
\begin{proof}
  We prove that (i) and (ii) hold with respect to $\CNF^P_k(\text{tw})$.
  Because  $P(\FALL(n))$ is the
  complete graph on $n$ vertices, it has treewidth $n-1$~\cite{Bodlaender98,KloksBodlaender92}. 
  Hence,
   $\FALL(n)\notin \CNF^P_k(\text{tw})$ for all $n\geq k+2$. This proves~(i).
  Because $P(\FCYC(n))$ is a cycle of length $n$, it has treewidth 2~\cite{Bodlaender98,KloksBodlaender92}.
Hence,  $\FCYC(n)\in \CNF^P_2(\text{tw})$.  This proves~(ii).
  \end{proof}

\begin{proposition}\label{pro:tw2}
  For every $k\geq 2$,
  $\CNF^I_k(\text{tw})$ is incomparable with $\BAC$
  and $\DPS$.
\end{proposition}

\begin{proof}
 We prove that (i) and (ii) hold with respect to $\CNF^I_k(\text{tw})$.
Because  $I(\FALL(n))$ is a
  complete bipartite graph with partition classes of size $n$ and $2^n$, respectively, it has treewidth $n$~\cite{Bodlaender98,KloksBodlaender92}. 
  Hence,
   $\FALL(n)\notin \CNF^I_k(\text{tw})$ for all $n\geq k+1$. This proves (i).
  Because $I(\FCYC(n))$ is a cycle of length $2n$, it has treewidth 2~\cite{Bodlaender98,KloksBodlaender92}.
  Hence,
  $\FCYC(n)\in \CNF^I_2(\text{tw})$.  This proves~(ii).
\end{proof}

\begin{proposition}\label{pro:cw}
For every $k\geq 4$,  $\CNF^D_k(\text{cw})$ is incomparable with $\BAC$  and $\DPS$. 
\end{proposition} 
\begin{proof}
  First we show that $\BAC \setminus \CNF^D_k(\text{cw})$
  contains formulas with an arbitrary large number of variables.  
  For all $n\geq 1$, Brandst{\"a}dt and
  Lozin~\cite{BrandstadtLozin03} showed that there is a bipartite
  permutation graph $G(n)$ with clique-width $n$.  We do not need the
  definition of a bipartite permutation graph; it suffices to know that
  bipartite permutation graphs are 
 chordal bipartite~\cite{Spinrad03}.  
 
 Let $G'(n)=(U_n\cup W_n,E_n)$ denote the graph
  obtained from $G(n)$ by deleting twin vertices as long as possible;
  two vertices are {\it twins} if they have exactly the same neighbors.  The
  deletion of twins does not change the clique-width of a
  graph~\cite{CourcelleOlariu00}. Hence, $G'(n)$ has clique-width~$n$.
  It is well known and easy to see that the clique-width of a bipartite
  graph with partition classes of size $r$ and $s$, respectively, is not greater than $\min(r,s)+2$. Hence
  $\Card{U_n} \geq n-2$. Because we only deleted vertices, $G'(n)$ is also chordal bipartite. 
   
  Let $F(n)=\SB N(w) \SM w\in W_n
  \SE$ where $N(w)$ denotes the set of neighbors of $w$ in
  $G'(n)$. Then $G'(n)$ is the incidence graph of $F(n)$, 
  because $G'(n)$ has no twins. Hence $F(n)\in
  \BAC$ follows from Proposition~\ref{p-three}.  Recall that the clique-width of $G'(n)=I(F(n))$ is $n$ and 
  that $\Card{U_n}\geq n-2$.
  Since all clauses of $F(n)$ are positive, $I(F(n))$ and $D(F(n))$ have
  the same clique-width.  
   We conclude that $F(n)$ is a formula on at least
  $n-2$ variables that belongs to
  $\BAC\setminus \CNF^D_k(\text{cw})$ for $n\geq k+1$.
    
For the converse direction we observe that $D(\FCYC(n))$ is
an oriented cycle and clearly has clique-width at most~$4$. This means
that $D(\FCYC(n))\in \CNF^D_4($cw$)$.  By Lemma~\ref{lem:fcyc-fallcyc},
we have that $D(\FCYC(n))\notin \DPS$ for all $n\geq 3$.  We then
conclude that $\CNF^D_4($cw$) \setminus \DPS$ contains $D(\FCYC(n))$ for
all $n\geq 3$.  We are left to apply Proposition~\ref{pro:chain} to
complete the proof of Proposition~\ref{pro:cw}.
\end{proof} 

\medskip
Results similar to Propositions~\ref{pro:tw}--\ref{pro:cw} also hold for
the graph invariants branchwidth and rank-width, since a
class of graphs has bounded branchwidth if and only if it has bounded
treewidth~\cite{Bodlaender98}, and a class of directed graphs has
bounded rank-width if and only if it has bounded
clique-width~\cite{GanianHlinenyObdrzalek10}.

\section{Parameterized Complexity}\label{s-parameterized}

We study the complexity of SAT for formulas that are ``almost'' $\beta$-acyclic.
We define what it means to be almost $\beta$-acyclic in two different ways.
We base the distance measure on the notion of a strong backdoor set in Section~\ref{sec:backdoors}, and on the notion of 
$\beta$-hypertree width in Section~\ref{sec:beta-htw}.
We start with a short introduction into Parameterized Complexity and refer to
other sources~\cite{DowneyFellows99,FlumGrohe06} for an in-depth
treatment. 

A parameterized problem can be considered as a set of pairs
$(I,k)$, the instances, where $I$ is the main part and $k$ is the
parameter. The parameter is usually a non-negative integer.
The complexity class \XP\ consists of parameterized decision problems $\Pi$ such that for each instance $(I,k)$ it can be decided in 
$f(k)|I|^{g(k)}$ time whether $(I,k)\in \Pi$, where $f$ and $g$ are computable functions depending only on the parameter $k$, and $|I|$ denotes the size of $I$. So \XP\ consists 
of parameterized decision problems which can be solved in polynomial
time if the parameter is a constant.
A parameterized decision problem is \emph{fixed-parameter tractable} if
there exists a computable function $f$ such that instances $(I,k)$ of
size $n$ can be decided in time $f(k)n^{O(1)}$. The class $\FPT$ denotes
the class of all fixed-parameter tractable decision problems.

Parameterized complexity offers a completeness theory, similar to the
theory of $\NP$-completeness, that allows the accumulation of strong
theoretical evidence that some parameterized problems are not
fixed-parameter tractable. This theory is based on a
hierarchy of complexity classes $\FPT \subseteq \W[1] \subseteq \W[2]\subseteq \ldots 
\subseteq \XP$.  
Each class $\W[i]$ contains all parameterized decision problems that can be reduced to
a certain fixed parameterized decision problem under
\emph{parameterized reductions}.
These are many-to-one reductions where the parameter for one problem maps
into the parameter for the other. More specifically, problem $L$ reduces
to problem $L'$ if there is a mapping $R$ from instances of $L$ to
instances of $L'$ such that (i)~$(I,k)$ is a yes-instance of $L$ if and
only if $(I',k')=R(I, k)$ is a yes-instance of~$L'$, (ii)~$k'=g(k)$ for
a computable function $g$, and (iii)~$R$ can be computed in time
$f(k)n^{O(1)}$ where $f$ is a computable function and $n$ denotes the
size of $(I,k)$.  The class $\W[1]$ is
considered as the parameterized analog to \NP.

\subsection{Strong Backdoor Sets}\label{sec:backdoors}

Let $\CCC$ be a class of CNF formulas. Consider a CNF formula $F$
together with a set of variables $B \subseteq \var(F)$. We say that $B$
is a \emph{strong backdoor set} of $F$ with respect to 
$\CCC$ if for all truth assignments $\tau : B \rightarrow \{0,1\}$ we
have $F[\tau] \in \CCC$.  In that case we also say that $B$ is a
\emph{strong $\CCC$-backdoor set}.  For every CNF formula $F$ and every
set $B \subseteq \var(F)$ it holds that $F$ is satisfiable if and only
if $F[\tau]$ is satisfiable for at least one truth assignment $\tau: B
\rightarrow \{0,1\}$.  Thus, if $B$ is a strong $\CCC$-backdoor set of
$F$, then determining whether $F$ is satisfiable reduces to the
{\sc Satisfiability} problem for at most $2^{|B|}$ reduced CNF formulas
$F[\tau]\in\CCC$.

Now consider a strictly or permissively tractable class $\CCC$ of CNF
formulas. Then, if we
have found a strong $\CCC$-backdoor set of $F$ of size $k$, deciding
the satisfiability of $F$ is fixed-parameter tractable for parameter
$k$. Hence, the key question is whether we can find a strong backdoor
set of size at most $k$ if it exists.  To study this question, we
consider the following parameterized problem; note that this problem
belongs to \XP\ for every fixed strictly tractable class~$\CCC$.
\begin{quote}
  \textsc{Strong $\CCC$-Backdoor}
    
  \emph{Instance:} A formula $F$ and an integer $k > 0$. 
  
  \emph{Parameter:} The integer $k$.
  
  \emph{Question:} Does $F$ have a strong $\CCC$-backdoor set of size
  at most $k$?
\end{quote} 
It is known that \textsc{Strong $\CCC$-Backdoor} is fixed-parameter
tractable for the class $\CCC$ of Horn formulas and for the class $\CCC$
of 2CNF formulas~\cite{NishimuraRagdeSzeider04-informal}.    
Contrary to these
results, we show that \textsc{Strong $\BAC$-Backdoor} is $\W[2]$-hard.

\begin{theorem}\label{t-w2}
  The problem \textsc{Strong $\BAC$-Backdoor} is $\W[2]$-hard.
\end{theorem}
\begin{proof}
  Let $\SSS$ be a family of finite sets $S_1,\dotso,S_m$.  Then a subset
  $R\subseteq \bigcup_{i=1}^mS_i$ is called a \emph{hitting set} of
  $\SSS$ if $R \cap S_i \neq \emptyset$ for $i=1,\ldots,m$.  The
  \textsc{Hitting Set} problem is defined as follows.
\begin{quote}
    \textsc{Hitting Set}
    
    \emph{Instance:} A family $\SSS$ of finite sets $S_1,\dotso,S_m$ and
    an integer $k > 0$.
    
    \emph{Parameter:} The integer $k$.  

    \emph{Question:} Does $\SSS$ have a hitting set of size at most $k$?
   
  \end{quote}
  It is well known that \textsc{Hitting Set} is
  $\W[2]$-complete~\cite{DowneyFellows99}.  We reduce from this problem
  to prove the theorem.

  Let $\SSS=\SB S_1,\dotso,S_m \SE$ and $k$ be an instance of
  \textsc{Hitting Set}.  We write $V(\SSS)=\bigcup_{i=1}^mS_i$ and
  construct a formula $F$ as follows.  For each $s\in V(\SSS)$ we introduce a
  variable $x_s$, and we write $X=\SB x_s \SM s\in V(\SSS) \SE$.  For
  each $S_i$ we introduce two variables $h^1_i$ and $h^2_i$.  Then, for
  every $1 \leq i \leq m$, the formula $F$ contains three clauses $C_i,
  C_i^1,$ and $C_i^2$ such that:
  \begin{itemize}
  \item [$\bullet$] $C_i=\SB h_{i}^1, h_{i}^2 \SE$;
  \item [$\bullet$] $C_i^1=\SB h_{i}^1\SE \cup \SB x_s \SM s \in S_i \SE \cup \SB
    \ol{x}_s \SM s \in V(\SSS) \setminus S_i) \SE$;
  \item [$\bullet$] $C_i^2=\SB h_{i}^2\SE \cup \SB \ol{x}_s \SM s \in V(\SSS) \SE$.
\end{itemize}

We need the following claims. The first claim characterizes the induced
cycles in $I(F)$ with length at least 6. We need it to prove the second
claim.

\bigskip\noindent \emph{Claim 1.} Let $D$ be an induced cycle in
$I(F)$. Then $|V(D)|\geq 6$ if and only if
$V(D)=\{h_i^1,h_i^2,x_s,C_i,C_i^1,C_i^2\}$ for some $1\leq i\leq m$ and
$s\in V(\SSS)$.
\par\bigskip\noindent
We prove Claim 1 as follows.  Suppose that $D$ is an induced cycle in $I(F)$
with $|V(D)|\geq 6$. By construction, $D$ contains at least one vertex
from $X$.  Because any two vertices in $X$ have exactly the same
neighbors in $I(F)$, $D$ contains at most one vertex from $X$.  Hence,
$D$ contains exactly one vertex from $X$, let $x_s$ be this vertex.  Let
$C^j_i$ and $C^{j'}_{i'}$ be the two neighbors of $x_s$ on $D$.  Because
$x_s$ is the only of $D$ that belongs to $X$, we find that $h^j_i$ and
$h^{j'}_{i'}$ belong to $D$. By our construction, $C_i$ and $C_{i'}$
then belong to $D$ as well.  If $C_i\neq C_{i'}$, then $D$ contains at
least two vertices from $X$, which is not possible. Hence $C_i=C_{i'}$,
as desired.  The reverse implication is trivial, and Claim~1 is proven.

\bigskip
\noindent
\emph{Claim 2.} Let $B$ be a strong $\BAC$-backdoor set that contains
variable $h_i^j$.  Then, for any $s^*\in S_i$, the set $(B\backslash
\{h_i^j\})\cup \{x_{s^*}\}$ is a strong $\BAC$-backdoor set.
 
\bigskip
\noindent
We prove Claim 2 as follows. Let $s^*\in S_i$ and define
$B'=(B\backslash \{h_i^j\})\cup \{x_{s^*}\}$.  Suppose that $B'$ is not a
strong $\BAC$-backdoor set. Then there is a truth assignment $\tau:
B'\to \{0,1\}$ with $F[\tau]\notin \BAC$. This means that $I(F[\tau])$
contains an induced cycle $D$ with $|V(D)|\geq 6$.  Because $B$ is a
strong $\BAC$-backdoor set, $h_i^j$ must belong to $V(D)$. We apply
Claim 1 and obtain $V(D)=\{h_i^1,h_i^2,x_s,C_i,C_i^1,C_i^2\}$ for some
$x_s\in X$. Suppose $\tau(x_{s^*})=1$. Then $C_i^1\notin F[\tau]$.
Hence $\tau(x_{s^*})=0$, but then $C_i^2\notin F[\tau]$.  This
contradiction proves Claim 2.

We are ready to prove the claim that $\SSS$ has a hitting set of size at
most $k$ if and only if $F$ has a strong $\BAC$-backdoor set of size at
most $k$.

Suppose that $\SSS$ has a hitting set $R$ of size at most $k$.  We claim that
$B=\SB x_s \SM s\in R \SE$ is a strong $\BAC$-backdoor set of $F$.
Suppose not. Then there is a truth assignment $\tau$ with $F[\tau]\notin
\BAC$.  This means that $I(F[\tau])$ contains an induced cycle $D$ with
$|V(D)|\geq 6$.  By Claim 1, we obtain
$V(D)=\{h_i^1,h_i^2,x_s,C_i,C_i^1,C_i^2\}$ for some $1\leq i\leq m$ and
$s\in S$.  Because $C_i^1,C_i^2$ are in $I(F[\tau])$, we find that
$R\cap S_i=\emptyset$. This is not possible, because $R$ is a hitting
set of $\SSS$.

Conversely, suppose that $F$ has a strong $\BAC$-backdoor set $B$ of size at
most $k$.  By Claim~2, we may without loss of generality assume that $B
\subseteq X$. We claim that $R=\SB s \SM x_s\in B\SE$ is a hitting set
of $\SSS$.  Suppose not. Then $R\cap S_i=\emptyset$ for some $1\leq
i\leq m$.  This means that $B$ contains no vertex from $\SB x_s \SM s\in
S_i\SE$. Let $\tau: B\to \{0,1\}$ be the truth assignment with
$\tau(x_s)=1$ for all $x_s\in B$.  Then $C^1_i$ and $C^2_i$ are in
$F[\tau]$.  Let $s\in S_i$. Then the cycle $D$ with
$V(D)=\{h^1_i,h^2_i,x_s,C_i,C^1_i,C^2_i\}$ is an induced 6-vertex cycle
in $I(F[\tau])$. This means that $F[\tau]\notin \BAC$, which is not
possible. Hence, we have proven Theorem~\ref{t-w2}.
\end{proof}

We finish Section~\ref{sec:backdoors} by considering another type of 
backdoor sets.
Let $F$ be a formula and let $B \subseteq \var(F)$ be a set of
variables. 
Recall that  $F-B$ denotes the formula obtained from $F$ after
removing all literals $x$ and $\ol{x}$ with $x \in B$ from the clauses
in $F$. We call $B$ a \emph{deletion backdoor set} with respect to 
a class $\CCC$ if $F-B \in \CCC$.

Deletion $\CCC$-backdoor sets can be seen as a relaxation of strong
$\CCC$-backdoor sets if the class $\CCC$ is \emph{clause-induced},
i.e., if for every $F\in \CCC$ and $F' \subseteq F$, we have $F' \in
\CCC$.  In that case every deletion $\CCC$-backdoor set $B$ is also a
strong $\CCC$-backdoor set.  This is well
known~\cite{NishimuraRagdeSzeider07} and can easily be seen as
follows. Let $\tau: B\to \{0,1\}$ be a truth assignment. Then by
definition $F[\tau]\subseteq F-B$. Because $B$ is a deletion
$\CCC$-backdoor set, $F-B\in \CCC$. Because $\CCC$ is clause-induced and
$F[\tau]\subseteq F-B$, this means that $F[\tau]\in \CCC$, as required.

Now let $\CCC$ be a clause-induced class. Let $B$ be a smallest
deletion $\CCC$-backdoor set and let $B'$ be a smallest strong
$\CCC$-backdoor set. Then, from the above, we deduce $|B'|\leq |B|$. The
following example shows that $|B|-|B'|$ can be arbitrarily large 
for $\CCC=\BAC$, 
which is obviously clause-induced.
Let $F$ be the formula with
$\var(F)=\{x_1,\ldots,x_p,y_1,\ldots,y_p,z_1,\ldots,z_p\}$ for some
$p\geq 1$ and clauses 
\[\begin{array}{lcl}
C_1 &= &\{x_1,\ldots,x_p,y_1,\ldots,y_p\},\\
C_2 &= &\{\ol{y}_1,\ldots,\ol{y}_p,z_1,\ldots,z_p\},\\
C_3 &= &\{x_1,\ldots,x_p,z_1,\ldots,z_p\}.
\end{array}
\]
Then
$B=\{y_1\}$ is a smallest strong $\BAC$-backdoor set.  However, a
smallest deletion $\BAC$-backdoor set must contain at least $p$
variables.

Analogously to the \textsc{Strong $\CCC$-Backdoor} problem we define the
following problem, where $\CCC$ is a
fixed clause-induced class.
\begin{quote}
  \textsc{Deletion $\CCC$-Backdoor}\nopagebreak
    
  \emph{Instance:} A formula $F$ and an integer $k > 0$. 
 
  \emph{Parameter:} The integer $k$.   
  
  \emph{Question:} Does $F$ have a deletion $\CCC$-backdoor set of size
  at most $k$?
\end{quote}
Determining the parameterized complexity of \textsc{Deletion
  $\BAC$-Backdoor} is interesting, especially in the light of our
$\W[2]$-hardness result for \textsc{Strong $\BAC$-Backdoor}. In other
words, is the problem of deciding whether a graph can be modified into a
chordal bipartite graph by deleting at most $k$ vertices fixed-parameter
tractable in $k$?  Marx~\cite{Marx10} showed that the version of this
problem in which the modified graph is required to be chordal instead of
chordal bipartite is fixed-parameter tractable.

\subsection{$\beta$-Hypertree Width}\label{sec:beta-htw}

The hypergraph invariant \emph{hypertree width} was introduced by Gottlob,
Leone, and Scarcello~\cite{GottlobLeoneScarcello02}. It is defined via
the notion of a \emph{hypertree decomposition} of a hypergraph~$H$,
which is a triple $\TTT=(T,\kappa,\lambda)$ where $T$ is a rooted tree
and $\chi$ and $\lambda$ are labelling functions with $\chi(t) \subseteq
V(H)$ and $\lambda(t) \subseteq E(H)$, respectively, for every $t \in
V(T)$, such that the following conditions hold:

\begin{enumerate}
\item For every $e \in E(H)$ there is a $t \in V(T)$ such that $e
  \subseteq \chi(t)$.
\item For every $v \in V(H)$, the set $\SB t \in V(T) \SM v \in \chi(t)
  \SE$ induces a connected subtree of $T$.
\item For every $t \in V(T)$, it holds that $\chi(t) \subseteq
  \bigcup_{e \in \lambda(t)}e$.
\item For every $t \in V(T)$, if a vertex $v$ occurs in some
  hyperedge $e \in \lambda(t)$ and if $v\in
  \chi(t')$ for some node $t'$ in the subtree below $t$, then 
  $v\in \chi(t)$.
\end{enumerate}

\noindent 
The \emph{width} of a hypertree decomposition
$(T,\chi,\lambda)$ is $\max\SB |\lambda(t)| \SM t \in V(T) \SE$. The
\emph{hypertree width}, denoted $\hw(H)$, of a hypergraph $H$ is the
minimum width over all its hypertree decompositions. Many $\NP$-hard
problems such as CSP or Boolean database queries can be solved in
polynomial time for instances with associated hypergraphs of bounded
hypertree width~\cite{GottlobLeoneScarcello01a}.
  
Gottlob and Pichler \cite{GottlobPichler04} defined $\beta$-hypertree
width as a ``hereditary variant'' of hypertree width. The
\emph{$\beta$-hypertree width}, denoted $\bhw(H)$, of a hypergraph $H$
is defined as the maximum hypertree width over all partial hypergraphs
$H'$ of $H$.  Using the fact that $\alpha$-acyclic hypergraphs are
exactly the hypergraphs of hypertree
width~$1$~\cite{GottlobLeoneScarcello02}, one deduces that the
hypergraphs of $\beta$-hypertree width 1 are exactly the $\beta$-acyclic
hypergraphs. Unfortunately, the complexity of determining the
$\beta$-hypertree width of a hypergraph is not
known~\cite{GottlobPichler04}.  However, we show the following. Here, a
\emph{$\beta$-hypertree decomposition} of width $k$ of a hypergraph $H$
is an oracle that produces for every partial hypergraph $H'$ of $H$ a
hypertree decomposition of width at most~$k$.

\begin{theorem}\label{the:W[1]-beta}
  {\sc SAT}, parameterized by an upper bound $k$ on the $\beta$-hypertree
  width of a CNF formula $F$, is $\W[1]$-hard even if a
  $\beta$-hypertree decomposition of width $k$ for $H(F)$ is given.
\end{theorem}
\begin{proof}
  A \emph{clique} in a graph is a subset of vertices that are mutually
  adjacent.  A $k$\hy partite graph is \emph{balanced} if its $k$
  partition classes are of the same size.  A \emph{partitioned clique}
  of a balanced $k$\hy partite graph $G=(V_1,\ldots,V_k,E)$ is a clique
  $K$ with $|K\cap V_i|=1$ for $i=1\ldots,k$.  We devise a parameterized
  reduction from the following problem, which is
  $\W[1]$-complete~\cite{Pietrzak03}.
\begin{quote}
  \textsc{Partitioned Clique}\nopagebreak
  
  \emph{Instance:} A balanced $k$\hy partite graph $G=(V_1,\ldots,V_k,E)$.
  
  \emph{Parameter:} The integer $k$. 
  
  \emph{Question:} Does $G$ have a partitioned clique?
\end{quote}
Before we describe the reduction we introduce some auxiliary concepts.
For any three variables $z,x_1,x_2$, let $F(z,x_1,x_2)$ denote the
formula consisting of the clauses 
$$\{z,x_1,\ol{x_2}\},           
\{z,\ol{x_1},x_2\},           
\{z,\ol{x_1},\ol{x_2}\},      
\{\ol{z},x_1,x_2\},          
\{\ol{z},\ol{x_1},\ol{x_2}\}.$$ 
This formula has exactly three  satisfying
assignments, corresponding to the vectors 000, 101, and 110.  Hence each
satisfying assignment sets at most one out of $x_1$ and $x_2$ to true,
and if one of them is set to true, then~$z$ is set to true as well
(``$z=x_1+x_2$''). Taking several instances of this formula we can build
a ``selection gadget.'' Let $x_1,\dots,x_m$ and $z_1,\dots,z_{m-1}$ be
variables. We define $F^{=1}(x_1,\dots,x_m;z_1,\dots,z_{m-1})$ as the
union of $F(z_1,x_1,x_2)$, $\bigcup_{i=2}^{m-1}
F(z_{i},z_{i-1},x_{i+1})$, and $\{\{z_{m-1}\}\}$.  Now each satisfying
assignment of this formula sets exactly one variable out of
$\{x_1,\dots,x_m\}$ to true, and, conversely, for each $1\leq i \leq m$
there exists a satisfying assignment that sets exactly $x_i$ to true and
all other variables from $\{x_1,\dots,x_m\}$ to false.

Now we describe the reduction.  Let $G=(V_1,\ldots,V_k)$ be a balanced
$k$\hy partite graph for $k\geq 2$.  We write
$V_i=\{v_1^i,\dots,v_n^i\}$. We construct a CNF formula $F$.  As the
variables of $F$ we take the vertices of~$G$ plus new variables $z_j^i$
for $1\leq i \leq k$ and $1\leq j \leq n-1$.  We put
$F=\bigcup_{i=0}^kF_i$ where the formulas $F_i$ are defined as follows:
$F_0$ contains for any $u \in V_i$ and $v \in V_j$ ($i \neq j$)
  with $uv \notin E$ the clause $C_{u,v} = \SB \ol{u}, \ol{v} \SE \cup
  \SB w \SM w \in (V_i \cup V_j) \setminus \SB u,v\SE \SE$;
for $i>0$ we define
  $F_i=F^{=1}(v_1^i,\dots,v_n^i;z_1^i,\dots,z_{n-1}^i)$.
To prove Theorem~\ref{the:W[1]-beta} it suffices to show the
following two claims.

\medskip
\noindent
\emph{Claim 1.} $\bhw(H(F))\leq k$.
 
\medskip
\noindent
We prove Claim 1 as follows.
First we show that that $\bhw(H(F_0))\leq k$. Let $H_0'$ be a partial
hypergraph of $H(F_0)$. Let~$I$ be the set of indices $1\leq i \leq k$
such that some hyperedge of $H_0'$ contains $V_i$. For each $i\in I$ we
choose a hyperedge $e_i$ of $H_0'$ that contains $V_i$.  The partial
hypergraph $H_0'$ admits a trivial hypertree decomposition
$(T_0,\chi_0,\lambda_0)$ of width at most $k$ with a single tree node
$t_0$ where $\chi_0(t_0)$ contains all vertices of $H_0'$ and
$\lambda_0(t_0)=\SB e_i \SM i\in I\SE$.  Second we observe that
$\bhw(H(F_i))=1$ for $1\leq i\leq k$: $H(F_i)$ is $\beta$-acyclic, and
$\beta$-acyclic hypergraphs have $\beta$-hypertree width~1.

Now let $H'$ be an arbitrarily chosen partial hypergraph of $H(F)$.  For
$i=0,\ldots,k$, we let $H'_i$ denote the (maximal) partial hypergraph of
$H'$ that is contained in $H(F_i)$.  We let
$\TTT_0=(T_0,\chi_0,\lambda_0)$ be a hypertree decomposition of width at
most $k$ of $H_0'$ as defined above.  For $i=1, \dots, k$ we let
$\TTT_i=(T_i,\chi_i,\lambda_i)$ be a hypertree decomposition of width
$1$ of $H_i'$.  We combine these $k+1$ hypertree decompositions to a
hypertree decomposition of width at most $k$ for $H'$.  We will do this
by adding the decompositions $\TTT_1,\dots,\TTT_k$ to $\TTT_0$ one by
one and without increasing the width of $\TTT_0$.

Let $\TTT_i^*=(T_i^*, \chi_i^*,\lambda_i^*)$ denote the hypertree
decomposition of width at most $k$ obtained from $\TTT_0$ by adding the
first $i$ hypertree decompositions.  For $i=0$ we let $\TTT_0^*=\TTT_0$.
For $i>0$ we proceed as follows.

First we consider the case where there is a hyperedge $e \in H'_0$ with
$V_{i+1} \subseteq e$.  Observe that there exists a node $t \in
V(T_{i}^*)$ with $e \subseteq \chi(t)$.  We define
$\TTT_{i+1}^*=(T_{i+1}^*, \chi_{i+1}^*,\lambda_{i+1}^*)$ as follows.  We
obtain $T_{i+1}^*$ from the disjoint union of $T_i^*$ and $T_{i+1}$ by
adding an edge between $t$ and the root of $T_{i+1}$.  As the root of
$T_{i+1}^*$ we choose the root of $T_i^*$.  We set
$\chi_{i+1}^*(t)=\chi_i^*(t)$ for every $t \in V(T_i^*)$, and
$\chi_{i+1}^*(t)=\chi_{i+1}(t) \cup V_{i+1}$ for every $t \in
V(T_{i+1})$; we set $\lambda_{i+1}^*(t)=\lambda_i^*(t)$ for every $t \in
V(T_i^*)$, and $\lambda_{i+1}^*(t)=\lambda_{i+1}(t) \cup \{e\}$ for
every $t \in V(T_{i+1})$ (hence $\Card{\lambda_{i+1}^*(t)}\leq
\max(2,k)=k$). Consequently $\TTT_{i+1}^*$ has width at most $k$.

It remains to consider the case where there is no hyperedge $e \in H'_0$
with $V_{i+1} \subseteq e$.  We define $\TTT_{i+1}^*$ as follows.  We
obtain $T_{i+1}^*$ from the disjoint union of $T_i^*$ and $T_{i+1}$ by
adding an edge between an arbitrary node $t\in V(T_i^*)$ and the root of
$T_{i+1}$.  As the root of $T_{i+1}^*$ we choose the root of $T_i^*$.
We set $\chi_{i+1}^*=\chi_i^* \cup \chi_{i+1}$ and
$\lambda_{i+1}^*=\lambda_i^* \cup \lambda_{i+1}$.  Clearly
$\TTT_{i+1}^*$ has width at most $k$. This completes the proof of  Claim~1.

\medskip
\noindent \emph{Claim 2.} $G$ has a partitioned clique if and only if $F$
is satisfiable.

\medskip
\noindent
To prove Claim 2 we first suppose that $G$ has a partitioned clique
$K$. We define a partial truth assignment $\tau:V\to \{0,1\}$ by setting
$\tau(v)=1$ for $v\in K$, and $\tau(v)=0$ for $v\notin K$. This partial
assignment satisfies $F_0$, and it is easy to extend $\tau$ to a
satisfying truth assignment of~$F$.  Conversely, suppose that $F$ has a
satisfying truth assignment $\tau$.  Because of the formulas $F_i$,
$1\leq i \leq k$, $\tau$ sets exactly one variable $v^i_{j_i}\in V_i$ to
true. Let $K=\{v^1_{j_1},\dots,v^k_{j_k}\}$.  The clauses in $F_0$
ensure that $v^i_{j_i}$ and $v^{i'}_{j_{i'}}$ are adjacent in $G$ for
each pair $1\leq i < i' \leq k$, hence $K$ is a partitioned clique
of~$G$.  This proves Claim~2. 
\end{proof}

We finish this section by showing some consequences of 
Theorem~\ref{the:W[1]-beta} with respect to the clique-width and rank-width of a formula.
By definition, 
the clique-width of a CNF formula is always bounded by its directed
clique-width. However, in general the directed clique-width can be much
higher than the undirected one. It is well known that SAT is fixed-parameter tractable for the parameter directed
clique-width~\cite{CourcelleMakowskyRotics01,FischerMakowskyRavve06}.
Fischer, Makowsky, and Ravve~\cite{FischerMakowskyRavve06} developed a
dynamic programming algorithm that counts the number of satisfying truth
assignments in linear time for CNF formulas of bounded directed
clique-width. They also conjectured that their method can be extended to
work for formulas of bounded (undirected) clique-width.  However, the
reduction in the proof of Theorem~\ref{the:W[1]-beta} shows that this is
not possible unless $\FPT=\W[1]$.

\begin{corollary}\label{cor:W[1]-clique}
  {\sc SAT}, parameterized by an upper bound $k$ on the clique-width of the
  incidence graph of a formula $F$, is $\W[1]$-hard even if a
  $k$-expression for $I(F)$ is given.
\end{corollary}
\begin{proof}
We use the same parameterized reduction as in the proof of Theorem~\ref{the:W[1]-beta}. Hence it remains to prove that the clique-width of the incidence graph of the formula $F$ in the proof of Theorem~\ref{the:W[1]-beta} is at most
$k'=O(k)$. 
In fact, we show that a $k+4$-expression for the incidence
graph of $F$ can be obtained in polynomial time. 

We start with the following claim. 
Let $n\geq 3$, and for $i=1,\ldots,k$, let $T_i$ be the tree with vertices
$C_1^i,\ldots,C_{n-1}^i$, $v_1^i,\ldots,v^i_n$, $z_1^i,\ldots,z^i_{n-1}$, and edges $C_1^iv_1^i$, $C_1^iv_2^i$, $C_1^iz_1^i$, and
$C_j^iv_{j+1}^i$, $C_j^iz_{j-1}^i$, $C_j^iz_j^i$ for $j=2,\ldots,n-1$.

\medskip
\noindent
{\it Claim 1.} Every $T_i$ allows a $5$-expression resulting in a labeling in which every $C^i_j$ has label
$d$, every $v^i_j$ has label $i$, $z^i_{n-1}$ has label $e$, whereas every other
$z^i_j$ has label $d$.

\medskip
\noindent
Let $1\leq i\leq k$.
We prove Claim 1 by induction on $n$. 
Let $n=3$. We get a desired $5$-expression of $T_i$ in the following way.
We introduce $v_1^i$ and $v_2^i$, each with label $i$.
Then we introduce $C_1^i$ with label $b$. We perform the operation 
$\eta_{b,i}$ resulting in edges between $C_1^i$ and $v_1^i,v_2^i$, 
respectively. We introduce $z_1^i$ with label $c$ and perform
the operation $\eta_{b,c}$ resulting in an edge between $C_1^i$ and
$z_1^i$. We perform the operation $\rho_{b\rightarrow d}$ resulting in a change of label of $C_1^i$ from $b$ to $d$. 
We introduce $C_2^i$ with label $b$ and perform the operation $\eta_{b,c}$
resulting in an edge between $C_2^i$ and $z_1^i$. We perform the operation
$\rho_{c\rightarrow d}$ resulting in a change of label of $z_1^i$ from $c$ to 
$d$. 
We introduce $v_3^i$ with label $c$ and perform the operation $\eta_{b,c}$
resulting in an edge between $C_2^i$ and $v_3^i$. We perform the operation
$\rho_{c\rightarrow i}$ resulting in a change of label of $v_3^i$ from $c$ to 
$i$. We introduce $z^i_2$ with label $e$ and perform the operation 
$\eta_{b,e}$ resulting in an edge between $C_2^i$ and $z^i_2$. Hence, we have
obtained $T_3$. What is left to do is to perform the operation $\rho_{b\rightarrow d}$ resulting in a change of label of $C_2^i$ from $b$ to $d$.

Let $n\geq 4$. Suppose that we have a labeling of $T_{i-1}$ as in the statement of the claim. Then we do as follows. We introduce $C_{n-1}^i$ with label $b$ and perform the operation $\eta_{b,e}$ resulting in an edge between $C_{n-1}^i$ and 
$z_{n-2}^i$. We perform the operation
$\rho_{e\rightarrow d}$ resulting in a change of label of $z_{n-2}^i$ from $e$ to 
$d$. We introduce $v_n^i$ with label $c$ and perform the operation $\eta_{b,c}$
resulting in an edge between $C_{n-1}^i$ and $v_n^i$. We perform the operation
$\rho_{c\rightarrow i}$ resulting in a change of label of $v_n^i$ from $c$ to 
$i$. We introduce $z_{n-1}^i$ with label $e$ and perform the operation 
$\eta_{b,e}$ resulting in an edge between $C_{n-1}^i$ and $z_{n-1}^i$. Hence, we have
obtained $T_n$. What is left to do is to perform the operation $\rho_{b\rightarrow d}$ resulting in a change of label of $C_{n-1}^i$ from $b$ to $d$.
This completes the proof of Claim 1.

\medskip
\noindent
Note that in the proof of Claim 1 we never performed an operation $\eta_{d,x}$ 
for some $x\in \{b,c,d,e,i\}$. Hence, we can consider the trees in order
$T_1,\ldots, T_k$ to obtain a $(k+4)$-expression for
their disjoint union where $v_1^i,\ldots,v_k^i$ are
the (only) vertices of label $i$ for $i=1,\ldots,k$.
Moreover, we may assume that all other vertices have label $d$ because we can
apply the operation $\rho_{e\rightarrow d}$ afterwards. 
For $s=1$ and $t=2$ we now introduce a new vertex $D_{s,t}$ with label
$b$ and perform the operations $\eta_{b,s}$, $\eta_{b,t}$ to connect
$D_{s,t}$ to every $v^s_i$ and every $v^t_j$, respectively. Afterwards
we perform the operation $\rho_{b\rightarrow d}$ resulting in a change
of label of $D_{s,t}$ from $b$ to $d$. In this way, we can add a
vertex $D_{s,t}$ for every other index pair $1\leq s < t \leq k$ as
well while using no new labels.  We call the resulting graph $I'$.

We now return to the incidence graph $I(F)$ of the formula $F$ in the
proof of Theorem~\ref{the:W[1]-beta}. Observe that $I(F)$ can be
obtained from $I'$ by adding a number of copies of the vertices
$C^i_j$ and $D_{s,t}$. This does not increase the clique-width of
$I'$ 
as explained in the proof of Proposition~\ref{pro:cw}.
Hence, the clique-width of $I(F)$ is at most $k+4$, as required.
This completes the proof of Corollary~\ref{cor:W[1]-clique}.  
\end{proof}

The already mentioned graph parameter rank-width 
was introduced by Oum and Seymour
\cite{OumSeymour06} for approximating the clique-width of graphs. A
certain structure that certifies that a graph has rank-width at most $k$
is called a rank-width decomposition of width $k$.  Similar to
clique-width, one can define the rank-width of a directed graph that
takes the orientation of edges into account.  The \emph{directed} (or
\emph{signed}) \emph{rank-width} of a CNF formula is the rank-width of
its directed incidence graph.  
Ganian, Hlin\v{e}n\'{y}, and
Obdr\v{z}\'{a}lek~\cite{GanianHlinenyObdrzalek10} developed an efficient
dynamic programming algorithm that counts in linear time the number of
satisfying assignments of a CNF formula of bounded \emph{directed}
rank-width.  Because bounded undirected rank-width implies bounded
undirected clique-width~\cite{OumSeymour06}, the following is a direct
consequence of Corollary~\ref{cor:W[1]-clique}.
\begin{corollary}\label{cor:W[1]-rank}
  {\sc SAT}, parameterized by an upper bound $k$ on the rank-width of the
  incidence graph of $F$, is $\W[1]$-hard even if a rank-decomposition
  of width $k$ for $I(F)$ is given.
\end{corollary}

\section{Conclusion}\label{s-con}
We have studied new classes of CNF formulas: the strictly tractable
class $\BAC$, the permissively tractable class $\DPS_\forall$, and the
hard-to-recognize class $\DPS$. Our results show that the classes are
incomparable with previously studied classes. Moreover, they establish an
interesting link between SAT and algorithmic graph theory: the formulas
in $\BAC$ are exactly the formulas whose incidence graphs belong to the
class of chordal bipartite graphs, a prominent and well-studied graph
class. 
It would be interesting to study systematically other classes of bipartite
graphs, e.g., the classes described by
Brandst{\"a}dt, Le and Spinrad~\cite{BrandstadtLeSpinrad99}, in order to
determine the complexity of SAT restricted to CNF formulas whose
incidence graphs belong to the class under consideration. 

We have also established hardness results for two natural strategies for
gradually extending $\BAC$: extensions via strong backdoor sets and
extensions via $\beta$-hypertree decompositions. The first extension is
fixed-parameter intractable because it is $\W[2]$-hard to find a strong
backdoor set. The second extension is fixed-parameter intractable
because SAT is $\W[1]$-hard when parameterized by an upper bound on the
$\beta$-hypertree width even if the $\beta$-hypertree decomposition is
provided.  It would be interesting to know whether SAT belongs to \XP\
for CNF formulas of bounded $\beta$-hypertree width, if a
$\beta$-hypertree decomposition is provided.

\end{document}